\documentclass[12pt]{article}
\usepackage{xcolor}
\usepackage{textcomp}
\usepackage{chet}
\usepackage{amssymb,amsmath}
\usepackage{color}

\newcommand{\be}{\begin{equation}}
\newcommand{\ee}{\end{equation}}
\newcommand{\beq}{\begin{eqnarray}}
\newcommand{\eeq}{\end{eqnarray}}

\usepackage{tikz}
\DeclareMathOperator{\sign}{sgn}
\usepackage{bbm}
\usepackage{dsfont}
\usepackage{bbold}
\usepackage{arydshln}

\renewcommand{\vec}{}

\newcommand{\bone}{\mathbbm{1}}
\newcommand{\bzero}{\mathbb{0}}
\usepackage{flip-acronyms} 
\usepackage{slashed}
\usepackage{tikzfeynman,contour}

\usepackage[numbers,sort&compress]{natbib}

\newcommand{\bm}[1]{\mathbf{#1}}

\newcommand{\cN}{\mathcal{N}}

\usepackage{amsthm}

\usepackage{upgreek}

\usepackage[compat=1.0.0]{tikz-feynman}

\title{Fermions and Supersymmetry \vspace{.2cm} \\ in Neural Network Field Theories}

\author{Samuel Frank$^{1,}$\email{frank.sam@northeastern.edu}, James Halverson$^{1,3,}$\email{j.halverson@northeastern.edu}, Anindita Maiti$^{4,}$\email{amaiti@perimeterinstitute.ca}, Fabian Ruehle$^{1,2,3,}$\email{f.ruehle@northeastern.edu}}

\affiliation{$^{1}$Department of Physics, Northeastern University, Boston, MA 02115 USA\\ \vspace{.1cm}$^{2}$Department of Mathematics, Northeastern University, Boston, MA 02115 USA\\ \vspace{.1cm}$^{3}$The NSF AI Institute for Artificial Intelligence and Fundamental Interactions\\ \vspace{.1cm}$^{4}$Perimeter Institute for Theoretical Physics, Waterloo, ON, Canada}

\abstract{We introduce fermionic neural network field theories  via Grassmann-valued neural networks. Free theories are obtained by a generalization of the Central Limit Theorem to Grassmann variables. This enables the realization of the free Dirac spinor at infinite width and a four fermion interaction at finite width. Yukawa couplings are introduced by breaking the statistical independence of the output weights for the fermionic and bosonic fields. A large class of interacting supersymmetric quantum mechanics and field theory models are introduced by super-affine transformations on the input that realize a superspace formalism.}

\begin{document}
\maketitle

\tableofcontents\newpage

\section{Introduction}
A theoretical machine learning result of Neal \cite{neal} from the 1990s established a connection between neural networks in certain infinite-width limits and Gaussian processes. This result has been generalized to many different neural network architectures since 2015, including deep networks in the infinite width limit, convolutional networks \cite{Novak2018BayesianCN, GarrigaAlonso2019DeepCN} in the infinite channel limit, attention layers \cite{hron2020infinite} (used in Transformers and language models) in the limit of infinite number of heads, and many others \cite{williams, Matthews2018GaussianPB, schoenholz2017correspondence, Jacot2018NeuralTK}. See \cite{yangTP1, yangTP2} for a unified perspective. Gaussian processes, being defined by Gaussian densities on function space, realize free field theories, thereby establishing a connection between neural networks and field theories \cite{Halverson_2021}.

One may explore this connection more systematically as a new approach to field theory. The essential mathematical data are the functional form of a neural network $\phi_\theta(x)$, known as the architecture, and a density $P(\theta)$ on its parameters $\theta$. Together, they define a partition function
\begin{equation}
    Z[J] = \mathbb{E}[e^{\,\int d^dx\;J(x) \phi(x)}] = \int d\theta \, P(\theta) e^{\,\int d^dx \;J(x)\, \phi_\theta(x)}, \label{eqn:Z}
\end{equation}
where the second equality is known as the parameter space expression for the partition function (to be contrasted with the ordinary Feynman path integral) and we made the $\theta$ subscript explicit to emphasize that $\phi_\theta$-fluctuations are induced by $\theta$-fluctuations. The usual $J$-derivatives yield expressions for correlation functions expressed in parameter space, which may sometimes be evaluated explicitly as, e.g., for the two-point functions in the ML literature \cite{williams}.  
Under common assumptions (see, e.g., \cite{Yaida2019NonGaussianPA}), the connected correlators $\kappa$ exhibit Central Limit Theorem scaling, 
\begin{equation}
    \kappa^{(2r)} \propto \frac{1}{N^{r-1}}\;.
\end{equation}
This shows that the theory becomes Gaussian as $N\to \infty$ (i.e., out of the higher cumulants only $\kappa_2$ survives). In that case, a second description of the theory becomes available via a Gaussian path integral; this is the Gaussian process correspondence mentioned above. Even if the action is not known or the theory is interacting, the parameter space expression \eqref{eqn:Z} may be used to compute correlators and study the theory, including via Monte Carlo sampling of randomly initialized neural networks.

Studies of this neural network field theory (NN-FT) correspondence have thus far focused on scalar theories.
A number of physics results (the NN$\to$FT direction) are known, e.g. the origin of symmetries \cite{maiti2021symmetryviaduality}, a systematic understanding of the origin of interactions in relation to the Central Limit Theorem \cite{Halverson_2021}, relation to Euclidean \emph{quantum} field theories via the Osterwalder-Schrader axioms \cite{halverson2021building}, a technique for reconstructing actions from connected correlators \cite{demirtas2023neural}, a realization of $\phi^4$ theory, a construction of conformal theories in the embedding formalism \cite{halverson2024conformal}, and a detailed connection to quantum mechanics and reflection positivity \cite{ferko2025qmnn}. A number of results in the ``for-ML" direction are also known: for example, perturbative methods, including Feynman diagrams, were leveraged in \citep{Halverson_2021, banta2023structures, Dyer2020AsymptoticsOW, erdmenger2021quantifying,Berman_2025} to treat non-Gaussian (finite $N$) effects in \citep{Yaida2019NonGaussianPA, antognini2019finite, dinan2023effectivetheorytransformersinitialization}. Various renormalization group flow schemes, such as (a) Wilsonian RG for Gaussian process error propagations \citep{Howard_2025} and scaling laws \citep{coppola2025renormalizationgroupdeepneural} when combined with NN-FT, (b) information-theoretic exact RG (Polchinski's scheme) for optimal transport \citep{Cotler:2022fze}, Bayesian inference \citep{berman2022inverse, Berman_2023}, and field theories \citep{howard2024bayesianrgflowneural}, and (c) other non-perturbative RG \citep{Erbin:2021kqf,erbin2023functional}, were employed to NNs. A subset of the authors showed  that global symmetries of NN-FTs can be engineered via symmetry invariances of NN parameter distributions in \citep{maiti2021symmetryviaduality}. Related pedagogical background on field theory approaches to ML may be found in the review \cite{ringel2025applications}, the textbook \cite{roberts2022principles}, or the TASI lectures \cite{halverson2024tasi}.

In this paper we introduce fermionic neural network field theories via Grassmann-valued neural networks. Such Grassmann-valued fields are the standard way to incorporate the anticommuting statistics of fermions in the path integral, and therefore we would like to have a similar construction in the neural network setting in order to extend the NN-FT correspondence. 

In analogy to the scalar NN-FT case, an essential question is whether a Gaussian limit exists for Grassmann NN-FTs, which we answer in the affirmative by a generalization of the Central Limit Theorem; see Section \ref{sec:CLT}.
 If a Grassmann network architecture is comprised of output weights $a_{ij}$ and post-activations $\rho_j$ (which themselves could be complicated deep neural networks),
 \begin{equation}
     \Psi_i(x) = \frac{1}{\sqrt{N}} \sum_{j=1}^N a_{ij} \rho_j(x),
 \end{equation}
 then either the case of Grassmann weights or Grassmann post-activations may yield a free Grassmann theory in the $N\to \infty$ limit; see Section \ref{sec:GrassmannNNFT}. These ideas are applied concretely to well-known cases in Section \ref{sec:applications}, including a realization of the free Dirac spinor, Yukawa couplings, and a four-fermion interaction. In Section \ref{sec: SUSY}, new classes of interacting supersymmetric theories are realized. In Section \ref{sec:conclusions} we conclude and summarize our results; the interested reader might skip there immediately for a detailed summary.

\bigskip
While finalizing this manuscript, another paper on fermionic NN-FT appeared \cite{OtherPaper}. Their approach differs significantly from ours in that their neural network is not an anticommuting object but rather complex matrix-valued, such that fermionic behavior only arises in free correlators at infinite width. In contrast, our approach leverages Grassmann-valued neural networks to obtain fermionic behavior, including interactions, at all widths.

\section{Grassmann Variables and the Central Limit Theorem \label{sec:CLT}}

This section introduces Grassmann-valued random variables and their statistics in certain
asymptotic limits. Among various asymptotic limits, we are interested in those where sums
over independently and identically drawn Grassmann random variables become draws from
Gaussian distributions, in the sense that all higher-order connected moments vanish. The origin of non-Gaussianities in distributions over Grassmann
random variables can then be studied as parametric deviations from these Gaussian limits.

To facilitate this direction, we derive a Central Limit Theorem for both discrete and continuous sets of Grassmann variables. We will prove in Section \ref{sec:discretegrassmann}:

\vspace{.5cm}
\noindent\begin{minipage}{1.0\textwidth}
  \centering
  \begin{minipage}{.9\textwidth}
    \textbf{Grassmann CLT:} a sum over $N$ i.i.d.\  Grassmann random variables is Gaussian distributed as $N \to \infty$ .
  \end{minipage}
\end{minipage}

\vspace{.6cm}
\noindent As in the real-valued case in \citep{Yaida2019NonGaussianPA,demirtas2023neural}, we will demonstrate parametric control of Grassmann non-Gaussianities by violating assumptions of the CLT, especially independence and $N\to \infty$.

\subsection{Discrete Case  \label{sec:discretegrassmann} }
Let us start our discussion with a scaled sum over $N$ randomly drawn discrete Grassmann-valued vectors $\vec{X_i} = (X_{i1}, \cdots , X_{id})$, where $i = 1, \cdots, N$ labels the statistical axis (i.e., the one summed over in the Central Limit Theorem), and each vector lives in a $d$-dimensional internal space. The weighted sum is taken along the statistical axis,
\begin{align}
    \psi_j = \frac{1}{\sqrt{N}} \sum_{i=1}^N  X_{ij},\quad j = 1 , \cdots , d\,. \label{eq:grassmann1}
\end{align}
Here, $\psi_j$ is the $j$-th element of a $d$-dimensional discrete Grassmann-valued vector.\footnote{\label{footnote:grassmann_subtlety}Throughout, by a Grassmann-valued vector $\psi_i$, we implicitly mean $\psi_i=a_{i} \odot \xi_i$ where the Hadamard product keeps track of the fact that we have a coefficient $a_i$ for each Grassmann variable, which takes values in some field, and the Grassmann variable $\xi_i$ gives anti-commuting statistics.  }
Random variables $X_{ij}$ are sampled from mean-zero distributions i.i.d.\  across the statistical axis:
\begin{align}
    &P(X) = \prod_{i=1}^N P(X_{i1},\dots,X_{id})\;, \\
    &P(X_{11},\dots,X_{1d}) = \ldots = P(X_{N1},\dots,X_{Nd})\;, \qquad \mathbb{E}[X_{ij}]=0 ~\forall~i,j\;, \label{eq:discretemeanfree}
\end{align}
yet statistically correlated along the internal axis,
\begin{align}
P(X_{i1},\cdots,X_{id}) \neq \prod_{j=1}^d P(X_{ij})\;.
\end{align}
Such correlations are necessary to ensure a nontrivial two-point function.
The Grassmann nature is manifest in elementwise anticommutation relations, 
\begin{align}
    \{X_{ij}, X_{kl} \} = 0~~\forall~i,j,k,l\;.
\end{align}
With this, one obtains $\{\psi_j, \psi_k \} = 0\,\, \forall j, k$.

\smallskip
Next, we wish to study the distribution of $\vec{\psi}:= (\psi_1,\cdots,\psi_d)$, with an emphasis on asymptotic limits for Gaussianity. To that end, we construct the cumulant generating function (CGF) of $\vec{\psi}$. This approach will be extrapolated to the continuous Grassmann random variable case in Section \ref{sec:contgrassmann}.  

Introduce a $d$-dimensional vector $\Vec{J}= (J_1, \cdots , J_d )$ as a Grassmann-valued source that anticommutes with $\vec{\psi}$ elementwise as $\{\psi_i, J_j \} = 0~\forall~i,j$. Then, let us define the cumulant generating function $W_\psi[J]$ as the logarithm of the partition function,
\begin{align} \label{eqn:discreteGrasmmannCGF}
    W_{\vec{\psi}}[\vec{J}] &= \log \mathbb{E}_{P(\vec{\psi})}\big[e^{ J_j \psi_j} \big]  = \log \mathbb{E}_{P(X)} \big[e^{\frac{1}{\sqrt{N}} \sum_{i} J_j X_{ij} }  \big] \notag \\
    &= \sum_{i=1}^N \log \mathbb{E}_{ P(X_{i1},\cdots,\,X_{id}) }\big[ e^{\frac{1}{\sqrt{N}}  J_j X_{ij} } \big]
    =  \sum_{i=1}^N W_{X_{i}}\bigg[\frac{\vec{J}}{\sqrt{N}}\bigg].
    \end{align}
We have used \eqref{eq:sumGrassmann} and \eqref{eq:dotGrassmann} to obtain the last equality. 
The CGF can also be expanded as a sum of connected moments (cumulants):
\begin{align}
W_{\Vec{\psi}}[\vec{J}]   &=
   \sum_{r=1}^{d}(-1)^{r(r-1)/2} \, \frac{J_{j_1}\cdots J_{j_r}}{r!} \kappa^{\vec{\psi}}_{j_1\cdots j_r}\nonumber \\
   &=  \sum_{i=1}^N W_{X_{i}}\bigg[\frac{\vec{J}}{\sqrt{N}}\bigg]=   \sum_{i=1}^{N}  \sum_{r=1}^{d}(-1)^{r(r-1)/2} \, \frac{J_{j_1}\cdots J_{j_r}}{r!N^{r/2}} \kappa^{X_{ij}}_{j_1\cdots j_r} \label{eq:cgf_expansion},
\end{align}
where
\begin{align}
    \kappa^{\vec{\psi}}_{j_1\cdots j_r} &:= \mathbb{E}^{(c)}_{P(\vec{\psi})}\big[\psi_{j_1} \cdots \psi_{j_r}\big] ,\nonumber \\
    \kappa^{X_{ij}}_{j_1\cdots j_r} &:= \mathbb{E}^{(c)}_{P(X_{i1},\cdots,\,X_{id})}\big[X_{ij_1} \cdots X_{ij_r}\big]
\end{align}
denote elements\footnote{$\kappa^{X_{ij}}_{j_1\cdots j_a \cdots j_b \cdots j_c \cdots  j_r} = {\sign(\pi)}\,\kappa^{X_{ij}}_{j_1\cdots j_c \cdots j_b \cdots j_a \cdots  j_r}$ for all $j$ due to antisymmetry, where $\pi$ is the permutation on the indices and  $\sign(\pi)$ is $+1$ for even permutations and $-1$ for odd.} in the rank-$r$ connected moments, with $\mathbb{E}^{(c)}$ labeling the connected parts of statistical moments.\footnote{{Unlike the cumulant expansion of the CGF for commuting variables, the Taylor series over $r$ in \eqref{eq:cgf_expansion} truncates at $r=d$ because $J_{j_1}\cdots J_{j_r}=0$ for $r>d$.}}
Equation \eqref{eq:cgf_expansion} implies a relation between the connected correlators of $\psi$ and the connected correlators of $X_{ij}$:
\begin{align}
    \kappa^{\vec{\psi}}_{j_1\cdots j_r} = \sum_{i=1}^{N} \frac{ \kappa^{X_{ij}}_{j_1\cdots j_r}}{N^{r/2}}\;,
\end{align}
which can be simplified via the identical-ness condition in \eqref{eq:discretemeanfree}:
\begin{align}
   \kappa^{\vec{\psi}}_{j_1\cdots j_r}= \frac{ \kappa^{X_{ij}}_{j_1\cdots j_r}}{N^{r/2 - 1}}\;,
\end{align}
demonstrating that the connected moments of $\psi$ of order $r>2$ vanish when $N \to \infty$.

We further need to show that the first and second connected correlators of $\vec{\psi}$ are finite. The first connected correlator vanishes due to mean-free assumptions in \eqref{eq:discretemeanfree}, and while the diagonal part of the second connected correlator vanishes, $\kappa^{\vec{\psi}}_{jj} = 0$, the off-diagonal part (which requires $d\geq 2$) is 
\begin{align}
    \kappa^{\vec{\psi}}_{j_1j_2} &= \mathbb{E}^{(c)}_{P(X_{i1},\cdots,\,X_{id})}[X_{ij_1}X_{ij_2}],
\end{align}
and is generally non-zero. If it is finite, which is case-dependent, then the vanishing of higher connected moments shows that $\vec{\psi}$ is Gaussian in the infinite width limit.

\subsubsection*{Non-Gaussianity from Statistical Independence Breaking}

We wish to systematically introduce non-Gaussianities in the distribution of $\vec{\psi}$. To do so, one may choose a finite value for the width $N$ so that non-Gaussianities are parametrically suppressed as $\mathcal{O}(N^{1-r/2})$ in the $r$-th connected moment. 

Alternatively, one may violate statistical independence of the variables $X_{ij}$ along the $i$ index by defining a joint distribution through a hyperparameter $\vec{\alpha} \in \mathbb{C}^q$ such that
\begin{align} \label{eqn:jointDist}
    P(X; \vec{\alpha}=\vec{0}) = \prod_{i=1}^N P(X_{i1},\cdots,X_{id})\;,
\end{align}
and any $\Vec{\alpha} \neq 0$ turns on statistical correlations. One can then expand $P(X; \vec{\alpha})$ as a Taylor series:
\begin{align}
    P(X; \vec{\alpha}) = P(X; \vec{\alpha}= \vec{0}) + \sum_{r=1}^{\infty}\sum_{s_1,\cdots, s_r = 1}^{q} \frac{\alpha_{s_1} \cdots \alpha_{s_r}}{r!}\partial_{\alpha_{s_1}}\cdots \partial_{\alpha_{s_r}} P(X; \vec{\alpha}) \Big|_{\vec{\alpha} = \vec{0}}\;,\label{eq:P_taylor}
\end{align}
where, for simplicity, we assume that $\alpha$-dependence enters only through Grassmann-even terms, so the Taylor coefficients in \eqref{eq:P_taylor} are $c$-valued.\footnote{By ``$c$-number'' or ``$c$-valued number'' we mean a commuting object; e.g., a real number or a product of an even number of anticommuting variables.} With these assumptions, the cumulant generating function inherits correlations among the random variables $X$ analogously to the $c$-valued case (see~\citep{demirtas2023neural}).

Nonetheless, we will still review the derivation with Grassmann variables inserted appropriately. We begin by introducing a popular trick below:
\begin{align} \label{eqn:policyGradThm}
    P \partial_{\alpha} \log P = \partial_{\alpha} P~~~~~~\partial_{\alpha}\mathbb{E}[\mathcal{O}] = \mathbb{E}[\mathcal{O} \partial_{\alpha} \log P],
\end{align}
for any $\alpha$-independent operator $\mathcal{O}$. Then, one can define an operator
\begin{align}
    \mathcal{P}_{r, \{ s_1, \cdots , s_r\}} := \frac{1}{P(\vec{X}; \vec{\alpha})} \partial_{\alpha_{s_1}}\cdots \partial_{\alpha_{s_r}} P(\vec{X}; \vec{\alpha})
\end{align}
that satisfies the following recursion relation:
\begin{align} \label{eqn:recursion}
    \mathcal{P}_{r+1, \{ s_1, \cdots , s_{r+1}\}} = \frac{1}{r + 1}\sum_{\gamma = 1}^{r + 1} ( \mathcal{P}_{1, s_{\gamma}} + \partial_{\alpha_{s_{\gamma}}} )\mathcal{P}_{r, \{ s_1, \cdots , s_{r+1}\}\backslash s_{\gamma}}.
\end{align}
With this notation, the CGF of $\vec{\psi}$ can be expanded as:
\begin{align}
    W_{\vec{\psi}}[\vec{J}] = \log \Bigg[e^{W_{\vec{\psi}, \vec{\alpha} = \vec{0}}[\vec{J}] } + \sum_{r=1}^{\infty}\sum_{s_1,\cdots, s_r = 1}^{q} \frac{\alpha_{s_1} \cdots \alpha_{s_r}}{r!}  \mathbb{E}_{P(\vec{X};\, \alpha=0)} \Big[e^{\frac{1}{\sqrt{N}}\sum_i\vec{J} \cdot \vec{X}_i}  \mathcal{P}_{r, \{ s_1, \cdots , s_{r}\}} \big|_{\vec{\alpha} = \vec{0}} \Big]  \Bigg],
\end{align}
where the free CGF at $\alpha=0$ is given in \eqref{eq:cgf_expansion}.
When all $|\alpha_i|$ are sufficiently small such that the $r=1$ term dominates, one can use the Taylor expansion $\log (1 + x) \approx x + \mathcal{O}(x^2)$ to show that
\begin{align} \label{eq:discreteCGFseriessum}
    W_{\vec{\psi}}[\vec{J}] = W_{\vec{\psi}, \vec{\alpha} = \vec{0}}[\vec{J}]  + \sum_{j=1}^q \frac{\alpha_j}{e^{ W_{\vec{\psi}, \vec{\alpha} = \vec{0}}[\vec{J}]  } }  \mathbb{E}_{P(\vec{X};\, \alpha=0)}\Big[e^{\frac{1}{\sqrt{N}}\sum_i \vec{J} \cdot \vec{X}_i } \mathcal{P}_{1, j } \big|_{\vec{\alpha} = \vec{0}} \Big]  + \mathcal{O}(\vec{\alpha}^2),
\end{align}
at leading order. Thus, in general higher-order connected moments will receive $\vec{\alpha}$-dependent non-vanishing contributions even as $N \to \infty$. 

When the hyperparameter $\vec{\alpha}$ is chosen as a Grassmann-valued vector instead, with $\{\alpha_i , \alpha_j \} = 0$, the CGF becomes 
\begin{align} \label{eqn:grassmannIBdiscrete}
    W_{\vec{\psi}}[\vec{J}] = \log \Bigg[e^{W_{\vec{\psi}, \vec{\alpha} = \vec{0}}[\vec{J}] } + \sum_{r=1}^{q}\sum_{s_1 \neq \cdots \neq s_r = 1}^{q} \frac{\alpha_{s_1} \cdots \alpha_{s_r}}{r!} \mathbb{E}_{P(\vec{X};\, \alpha=0)}\Big[e^{\frac{1}{\sqrt{N}}\sum_i \vec{J} \cdot \vec{X}_i } \mathcal{P}_{r, \{ s_1, \cdots , s_{r}\}} \big|_{\vec{\alpha} = \vec{0}} \Big]  \Bigg],
\end{align}
where the Grassmann nature of $\alpha_i$ facilitates the truncation of the sum over $r$ at $q$. The leading order corrections, i.e. at $\mathcal{O}(\vec{\alpha})$, are the same as in \eqref{eq:discreteCGFseriessum}. 

\subsection{Continuous Case \label{sec:contgrassmann}}
Having defined the Central Limit Theorem for discrete random Grassmann variables, let us address continuous Grassmann random variables. 

We start with a scaled sum over $N$ randomly drawn continuous Grassmann-valued functions $X_{i}(x)$ of $x \in \mathbb{R}^d$, $i=1,\cdots,N$:
\begin{align}
    \label{eqn:continuum CLT grassmann}
    \psi(x) = \frac{1}{\sqrt{N}}\sum_{i=1}^{N} X_{i}(x)\;.
\end{align}
By imposing the anticommutation relation $\{X_{i}(x),X_{j}(y) \}=0~\forall~x,y,i,j$, we obtain
\begin{align}
    \{\psi(x),\psi(y)\} =0~\forall~x,y\;.
\end{align}
Let us assume that all $X_{i}(x)$ are sampled independently from identical mean-zero distributions. 

Having introduced this setup, we are interested in studying the distribution over $\vec{\psi}(x)$
via the cumulant generating functional (CGF) approach. Let us introduce a Grassmann-valued continuous source current $J(x)$ that anticommutes with $\psi(x)$ as $\{\psi(x), J(y) \}=0~ \forall ~x,y$. Then, the CGF of $\psi(x)$ is 
\begin{align} \label{eqn:discreteGrasmmann1}
    W_{\Vec{\psi}}[\vec{J}] &= \log \mathbb{E}_{P(\vec{\psi})}\big[e^{ \int d^dx \, J(x) \psi(x)} \big]  = \log \mathbb{E}_{P(X)} \big[e^{\frac{1}{\sqrt{N}} \sum_{i=1}^N \int d^dx \, J(x) X_{i}(x) }  \big] \notag \\
    &= \sum_{i=1}^N \log \mathbb{E}_{ P(X_{i}) }\big[ e^{\frac{1}{\sqrt{N}}  \int d^dx\, J(x) X_{i}(x) } \big]
    =  \sum_{i=1}^N W_{X_{i}}\bigg[\frac{\vec{J}}{\sqrt{N}}\bigg],
    \end{align}
using \eqref{eq:sumGrassmanncont} and \eqref{eq:dotGrassmanncont} in the second to last equality. The connected correlators of $\psi(x)$ are therefore related to the connected correlators of $X_{i}(x)\,\forall \, x$ as
\begin{align}
& \sum_{r=1}^{\infty}(-1)^{r(r-1)/2} \int d^dx_1 \cdots d^dx_r \, \frac{J(x_1)\cdots J(x_r)}{r!} \kappa^{(r)}_{\vec{\psi}}(x_1,\cdots, x_r)\nonumber \\
   &=   \sum_{i=1}^{N}  \sum_{r=1}^{\infty}(-1)^{r(r-1)/2} \int d^dx_1 \cdots d^dx_r \, \frac{J(x_1)\cdots J(x_r)}{r!\,N^{r/2}} \kappa^{(r)}_{X_{i}}(x_1,\cdots, x_r) ,
\end{align}
where
\begin{align}
    \kappa^{(r)}_{\vec{\psi}}(x_1,\cdots, x_r) &:= \mathbb{E}^{(c)}_{P(\vec{\psi})}\big[\psi(x_1) \cdots \psi(x_r)\big] ,\nonumber \\
    \kappa^{(r)}_{X_{i}}(x_1,\cdots, x_r) &:= \mathbb{E}^{(c)}_{P(X_{i})}\big[X_{i}(x_1) \cdots X_{i}(x_r)\big]. 
\end{align}
Then, just as in the discrete case, the relation between the connected correlators of $\psi$ and the connected correlators of $X_i$ is found to be:
\begin{align}
   \kappa^{(r)}_{\vec{\psi}}(x_1,\cdots, x_r)  = \sum_{i=1}^{N} \frac{ \kappa^{(r)}_{X_{i}}(x_1,\cdots, x_r) }{N^{r/2}} = \frac{ \kappa^{(r)}_{X_{i}}(x_1,\cdots, x_r) }{N^{r/2 - 1}}\;,
\end{align}
where we once again see that the connected correlators of order $r>2$ vanish when $N \to \infty$. We can also prove that the first and second connected correlators are finite: $\kappa^{(1)}_{\psi}=0$ due to the mean-free assumption, and by the identical-ness assumption, the second connected correlator is:
 \begin{align}
     \kappa^{(2)}_{\psi}(x,y) = \mathbb{E}_{P(X_1)}[X_1(x)X_1(y)]\;,
 \end{align}
which can be chosen finite with appropriate $P(X_1)$.

\subsubsection*{Non-Gaussianity from Statistical Independence Breaking in the Continuous Case}
We aim to introduce non-Gaussianities into the distribution of $\vec{\psi}(x)$. Similar to the discrete case, we employ a hyperparameter $\vec{\alpha}\in\mathbb{C}^q$ such that the joint distribution is defined as
$$P\big(X(x);\,\vec{\alpha}=0\big)=\prod_{i=1}^N P\big(X_i(x)\big),$$
analogous to \eqref{eqn:jointDist}. Then, we can write the cumulant generating function \eqref{eqn:discreteGrasmmann1} in terms of $P\big(X(x);\,\vec{\alpha}\big)$:
 \begin{equation} \label{eqn:cgfIBCont}
     W_{\vec{\psi}}[\vec{J}]=\log\Big[\int D[X(x)]\,P\big(X(x);\, \vec{\alpha}\big)\;e^{\frac{1}{\sqrt{N}}\sum_{i=1}^N\int d^dx J(x) X_{i}(x)}\Big]\;,
 \end{equation}
 where
 \begin{equation}
   D[X(x)]:=\prod_{i=1}^N D[\vec{X}_i(x)]\notag
 \end{equation}
 is a functional integration measure over each $X_i(x)$.

 Then, using the same trick from the policy gradient theorem that was used in the discrete case, \eqref{eqn:policyGradThm}-\eqref{eqn:recursion}, the CGF \eqref{eqn:cgfIBCont} can be written as
 \begin{align}
     W_{\vec{\psi}}[\vec{J}]
     &=W_{\vec{\psi},\vec{\alpha}=0}[\vec{J}]+\sum_{j=1}^q\frac{\alpha_j}{e^{W_{\vec{\psi},\vec{\alpha}=0}[J]}}\mathbb{E}_{P(\vec{X};\, \alpha=0)}\Big[e^{\frac{1}{\sqrt{N}}\sum_{i=1}^N\int d^dx J(x)X_i(x)}\mathcal{P}_{1,j}\big|_{\vec{\alpha}=0}\Big]+\mathcal{O}(\vec{\alpha}^2),
 \end{align}
which is analogous to \eqref{eq:discreteCGFseriessum}. The case of Grassmann $\vec{\alpha}$ would be an extrapolation of  \eqref{eqn:grassmannIBdiscrete}.

\section{Grassmann Neural Network-Field Theories}
\label{sec:GrassmannNNFT}
We may utilize Grassmann-valued variables in neural networks to obtain Grassmann NN-FTs.
We will begin in the continuum, writing a Grassmann-valued neural network as
\begin{align}
    \Psi_i(x) = \frac{1}{\sqrt{N}} \sum_{j=1}^{N} \psi_{ij}(x)\;;
\end{align}
a normalized sum of Grassmann-valued neurons $\psi_{ij}$ that can induce Gaussianity according to the results of Section \ref{sec:CLT}. Relative to the continuum case \eqref{eqn:continuum CLT grassmann}, we have gained an extra index $i$, so instead of a single Grassmann field for every $x$, we allow a vector of Grassmann fields $\Psi_i(x)$ for every $x$, where $i=1,\dots,D$. Physically, this is motivated by the fact that Grassmann fields often carry a spinor index, which we will utilize when we realize the free Dirac spinor in Section \ref{sec:dirac_spinors}. Moreover, we will also see momentarily in a case of interest that it is required to have non-trivial higher-point correlation functions.

Though a variety of functional forms could be chosen for $\psi_{ij}(x)$, for simplicity we restrict ourselves to the case of a bias-less linear output layer acting on post-activations $\rho$:
\be 
\psi_{ij}(x) = a_{ij} \rho_j(x),
\ee 
where the post-activations $\rho_j$ depend on parameters and could be non-trivial functions appearing in a deep neural network. In this case, the network is
\be
\Psi_i(x) = \frac{1}{\sqrt{N}} \sum_{j=1}^{N} a_{ij} \rho_j(x)\;.
\ee 
A Grassmann-valued $\Psi$ may be obtained in two simple cases: either the $a_{ij}$ are Grassmann-valued and the $\rho_j$ are $c$-valued, or the $a_{ij}$ are $c$-valued and the $\rho_{j}$ are Grassmann-valued.

\subsection{Grassmann NN-FT from Grassmann Weights \label{sec:grassmannweightnnft}} We wish to study the case that the output weights $a_{ij}$ are Grassmann-valued and the post-activations $\rho$ are $c$-valued.  To allow for the possibility of Gaussianity from the CLT, we will work in the case of statistical independence on the axis $j$ in $a_{ij}$ where the large-$N$ sum would occur:
\begin{align}
    P(a) = \prod_{j=1}^{N}P(a_{1j},\cdots , a_{Dj})\;,
\end{align}
and also identical-ness; i.e., all of the distributions in the product are the same. We also assume that the $a_{ij}$ have mean zero and that $\mathbb{E}[a_{ij}a_{kj}] \neq 0~\forall i \neq k$, which is necessary to obtain non-trivial two-point functions.\footnote{Factorization of $P$ across the Grassmann index yields a vanishing two-point function.} All other network parameters, implicit in the post-activations $\rho$, are labeled collectively as $\theta$ and drawn i.i.d.\  across the $j$ index from distributions independent of $P(a)$.

Let us consider a vector of $D$ mutually anticommuting source currents $J_i(x)$, to define the CGF of $\Psi$ as the following:
\begin{align}
    &W_{\Psi}[J] = \log \mathbb{E}\Big[e^{\,\, \sum\limits_{i=1}^{D} \int d^dx\, J_i(x)\Psi_i(x)} \Big] \nonumber \\
    &\quad  = \sum_{j=1}^{N}\log \mathbb{E}_{P(a_{1j},\cdots,\,a_{Dj})\,P(\theta_j)}\Big[e^{\frac{1}{\sqrt{N}} \sum\limits_{i=1}^D \int d^dx \,   J_i(x) \, a_{ij} \rho_{j}(x) } \Big], \nonumber \\
    &\quad  = \sum_{j=1}^{N} \sum_{r=1}^{\infty} \sum_{i_1,\cdots, i_r=1}^{D} \frac{(-1)^{r(r-1)/2}}{r!N^{r/2}} \int d^dx_1 \cdots d^dx_r\, J_{i_1}(x_1)\cdots J_{i_r}(x_r)\,  \kappa^{(r)}_{\psi_{i_1j}, \cdots, \psi_{i_rj}}(x_1,\cdots, x_r) , \label{eq:grass_weight_cumu1}
\end{align}
where \begin{align}\kappa^{(r)}_{\psi_{i_1j}, \cdots, \,\psi_{i_rj}}(x_1,\cdots, x_r) = \mathbb{E}^{(c)}_{P(a_{1j},\cdots,\,a_{Dj})\,P(\theta_j)}\Big[\psi_{i_1j}(x_1)\cdots \psi_{i_rj}(x_r) \Big].\label{eq:grass_weight_cumu2}
\end{align}
The subscript $\psi_{i_1j}, \cdots, \psi_{i_rj}$ denotes the set of neurons over which the connected correlator is evaluated. Finally, the connected correlators of $\Psi$ are related to the connected correlators of the neurons:
\begin{align}
    \kappa^{(r)}_{\Psi}(x_1,\cdots,x_r)\big|_{i_1,\cdots,i_r} =  & ~ \frac{ \kappa^{(r)}_{\psi_{i_1}, \cdots,\, \psi_{i_r}}(x_1,\cdots, x_r) }{N^{r/2-1}}\;,\label{eq:cumulantGeneral}
\end{align}
where we have dropped the statistical $j$ index.

\subsection{Grassmann NN-FT from Grassmann Post-Activations} 

We now turn to the case that the output weights $a_{ij}$ are $c$-valued, but the post-activations $\rho$ are Grassmann-valued. We treat this by performing an expansion in $p$ Grassmann variables. Specifically, to define $\rho$, consider a linear superposition in a $2^p$-dimensional Grassmann space spanned by generators $\Theta_1,\cdots, \Theta_p$:
\begin{align} \label{eq:grassmannNeurons}
    \rho_j =  \sum_{k=1}^{p} \sum_{\substack{l_1< \cdots< l_k \\ \in (1,\cdots, p) }} \sigma_{j,\;l_1 < \cdots < l_k}(y^{1},\cdots, y^{N})\,\Theta_{l_1}\cdots \Theta_{l_k},
\end{align}
where $l_1,\cdots , l_k$ are ordered subsets and $y^1,\dots, y^N$ denotes preactivations to the final hidden layer neurons. In the case of a network with a single hidden layer and input $x$, these preactivations are
\begin{align}
    y_i = \sum_{j=1}^{d} W^0_{ij} x_j + b^0_i\;. 
\end{align}
We also allow nonlinear activation functions $\sigma_{j,l_1\dots l_k}$ to differ from one statistical node to another. For anticommuting $\Psi$, the sum over $k$ in \eqref{eq:grassmannNeurons} is restricted only to odd integers.

To allow for the possibility of Gaussianity from the CLT, we will repeat our assumptions of statistical independence and identicalness of $a_{ij}$ on the $j$ axis where the large-$N$ sum occurs, as well as extend it to the axis $i$:
\begin{align}
    P(a) = \prod_{i=1}^{D} \prod_{j=1}^{N} P(a_{ij}),
\end{align}
with $P(a)$ chosen such that $\mathbb{E}[a_{ij}]=0$. All other network parameters, implicit in $\rho$, are denoted $\theta$ and drawn i.i.d.\  across the $j$ index such that all nonlinearities $\sigma_{j,\,l_1\dots l_k}$ in the post-activation $\rho_j$ are functions of the same parameters $\theta_j$. We note that the $2$-pt function $\langle\rho_j(x) \rho_j(y)\rangle$ does not generally vanish for $p\geq 2$ and distinct points $x\neq y$. 

Analogous to the previous example, we introduce a vector of $D$ mutually anticommuting source currents $J_i(x)$ to define the CGF of this network, 
\begin{align}
W_{\Psi}[J]
&= \log \mathbb{E}\Big[e^{\, \sum_{i=1}^{D} \int d^dx\, J_i(x)\Psi_i(x)} \Big] \nonumber \\
&= \sum_{j=1}^{N} \log \mathbb{E}_{P(a_{\cdot j}, \theta_j)}
\Big[e^{\frac{1}{\sqrt{N}} \sum_{i=1}^{D} a_{ij} \int d^dx \, J_i(x)\, \rho_{j}(x)} \Big] \nonumber \\
&= \sum_{j=1}^{N} \sum_{r=1}^{\infty} \sum_{i_1,\ldots,i_r=1}^{D}\!\!
\frac{(-1)^{r(r-1)/2}}{r!\,N^{r/2}}
\int d^dx_1 \cdots d^dx_r\,
J_{i_1}(x_1)\cdots J_{i_r}(x_r)\,
\kappa^{(r)}_{\psi_{i_1 j},\ldots, \psi_{i_r j}}(x_1,\ldots, x_r)\;,
\label{eq:grassNNFTcumu1}
\end{align}
where \begin{align}
\kappa^{(r)}_{\psi_{i_1j},\dots,\psi_{i_r j}}(x_1,\cdots, x_r) = \mathbb{E}^{(c)}_{P(a_{\cdot j},\theta_j)}\big[\psi_{i_1j}(x_1)\cdots \psi_{i_rj}(x_r) \big]\;.   
\end{align}
Exploiting identical-ness of post-activation parameters $\theta_j$, the connected correlators of $\Psi$ are related to the connected correlators of neurons as
\begin{align}
    \kappa^{(r)}_{\Psi_i}(x_1,\cdots,x_r) =  & ~ \frac{ \kappa^{(r)}_{\psi_i }(x_1,\cdots, x_r) }{N^{r/2-1}},
\end{align}
where we have dropped the statistical index.

\section{Applications in Fermionic Field Theories}
\label{sec:applications}

Having developed some essential formalism for Grassmann NN-FTs and Gaussian limits thereof, we now turn to applications in conventional fermionic field theories: the free Dirac spinor and Dirac spinor interactions.

\subsection{Free Dirac Spinors}\label{sec:dirac_spinors}

As a concrete example of the formalism given in previous sections, we construct one instance of a Grassmann-valued NN architecture with action equal to free Dirac theory in $4$-dim Euclidean space. Historically, the task of defining fermions in Euclidean space posed fundamental challenges, mainly enforcing Osterwalder-Schrader reflection positivity to guarantee a well-defined Lorentzian theory, as was done in the case of scalar fields \citep{Osterwalder1973}. Several works, such as \citep{OSfermions, SchwingerFermions, ZUMINO1977369, Mehtafermions, Nicolai:1978vc, VANNIEUWENHUIZEN199629, Waldron_1998}, constructed Dirac and other fermion theories on $4$-dim Euclidean space. For example, \citep{OSfermions} requires a ``doubling'' of fermionic degrees of freedom on Euclidean space; as such, the Dirac field $\Psi$ and its Dirac adjoint $\bar{\Psi}$ are no longer related by Hermitian conjugation in the Euclidean theory. Later works, e.g. \citep{ZUMINO1977369, SchwingerFermions}, ensure Hermitian Dirac fields on Euclidean space; however, the number of degrees of freedom in each field is still doubled. 

The free Dirac NN-FT architecture presented in this section is engineered by inverting the Euclidean free Dirac action of van Nieuwenhuizen and Waldron \citep{VANNIEUWENHUIZEN199629}. Among other advantages, this approach preserves Hermiticity of the Euclidean Dirac action by supplementing the normal analytic continuation $t\to i\tau$ with a unitary rotation on spinor indices. In a follow-up work \citep{Waldron_1998}, Waldron gives a canonical formulation of the spinor Wick rotation; although the Euclidean Fock space includes negative-norm modes, OS reflection positivity is preserved, thereby ensuring a positive-definite Lorentzian Hilbert space.

While our free Dirac NN-FT is explicitly defined in $4$ spatial dimensions, following \citep{VANNIEUWENHUIZEN199629}, one may nontrivially extend this recipe to dimensions other than four.

We start with the Lagrangian of a free Dirac fermion on four-dimensional Euclidean space~\citep{VANNIEUWENHUIZEN199629},
\begin{equation}
    \mathcal{L}=\Psi^\dagger_E \gamma^5_E(\gamma^\mu_E\partial_\mu+m)\Psi_E\label{lagrangian}\;,
\end{equation}
where $\gamma^5_E$ and $\gamma^\mu_E$ are Hermitian matrices satisfying $\{\gamma^\mu_E, \gamma^5_E\}=0$ and $\gamma^5_E:=\gamma^1_E \gamma^2_E \gamma^3_E \gamma^4_E$. Additionally, it can be shown that $\gamma^\mu_E$ has the proper Clifford algebra in Euclidean space,
\begin{equation}
    \{\gamma^\mu_E, \gamma^\nu_E\}=2\delta^{\mu\nu}\label{eq:cliff}\;,
\end{equation}
where $\mu,\nu=1,\dots,4$.

The free Dirac NN-FT $\Psi_{\text{E}}$ is defined as an asymptotically wide single hidden layer fully-connected feedforward architecture with exponential activation functions and complex Grassmann-valued outputs\footnote{A complex-valued Grassmann variable $\theta$ is defined from two real Grassmann variables $\theta_1$ and $\theta_2$ by $\theta=\theta_1+i\theta_2$. Real Grassmann variables satisfy $\theta_1^*=\theta_1$.}:
\begin{align} \label{eq:DiracNNFT}
    \Psi_{\text{E}}(x) = \begin{pmatrix}
        \Psi_1(x) \\ \vdots \\ \Psi_D(x) 
    \end{pmatrix}. 
\end{align}
This NN-FT exists in a dimension $D$ compatible with a matrix representation of the Clifford algebra in $d=4$ spatial dimensions. The NN field $\Psi_i(x)$ and its Hermitian conjugate are given by:
\begin{align}
    \Psi_{i}(x)&:=\sqrt{\frac{\text{Vol}(B^4_\Lambda)}{N\sigma^2 (2\pi)^4}}\sum_{j=1}^D \sum_{k=1}^N \frac{1}{\sqrt{E_k}}\Big[(\mathcal{P}^+_k)_{ij}\alpha_{jk}-(\mathcal{P}^-_k)_{ij}\beta^*_{jk}\Big]\exp\Big[i\big(\sum_{\mu=1}^4 b_{k\mu}x_{\mu}+c_k\big)\Big]\;, \label{eq:PsiNewNew}
    \\[10 pt]
{\Psi}^\dagger_i(x)&:=\sqrt{\frac{\text{Vol}(B^4_\Lambda)}{N\sigma^2 (2\pi)^4}}\sum_{j=1}^D \sum_{k=1}^N \frac{1}{\sqrt{E_k}}\Big[(\mathcal{P}^+_k)_{ji}\alpha^*_{jk}-(\mathcal{P}^-_k)_{ji}\beta_{jk}\Big]\exp\Big[-i\big(\sum_{\mu=1}^4 b_{k\mu}x_{\mu}+c_k\big)\Big] ,\label{eq:PsiDaggerNewNew} \;
\end{align}
and are based on complex Grassmann-valued parameters $\alpha$, $\beta$ and their conjugates $\alpha^*$, $\beta^*$, in addition to real $c$-valued parameters $b$ and $c$. Collectively, we choose the following parameter densities:
\begin{align}
    P(\alpha, \alpha^*)&= \sigma^{2ND} \prod_{i=1}^D \prod_{j=1}^N \exp\big[  \alpha_{ij}\,\alpha^*_{ij}/\sigma^2\big]\label{P(alpha)}\;,\\
    P(\beta,\beta^*)&=\sigma^{2ND}\prod_{i=1}^D\prod_{j=1}^N \exp\big[\beta_{ij}\,\beta^*_{ij}/\sigma^2\big]\;\label{P(beta)},\\
    P(b)&=\prod_{j=1}^N P(\mathbf{b}_j)\;,\quad\; P(\mathbf{b}_j)=\text{Unif}(B^4_\Lambda)\;,\label{P(b)}\\[10 pt]
    P(c)&=\prod_{j=1}^N \text{Unif}([-\pi,\pi])\label{P(c)}\;,
\end{align}
involving hyperparameters $\sigma^2 \in \mathbb{R}_{>0}$\footnote{The prefactor $\sigma^{2ND}$ in equations \eqref{P(alpha)} and \eqref{P(beta)} is chosen such that $P(\alpha,\alpha^*)$ (and $P(\beta,\beta^*)$) are properly normalized distributions, in the sense of $\int (\prod_{j,k}d\alpha^*_{jk}\,d\alpha_{jk}) \,P(\alpha,\alpha^*)=1$.} and $\Lambda$, the radius of the 4-ball $B^4_\Lambda$. Note that $P(\alpha,\alpha^*)$ and $P(\beta,\beta^*)$ are mean-zero complex Grassmann-valued Gaussian distributions with the \emph{same} variance $\sigma^2$, but the $\alpha,\alpha^*$ parameters are independent from the $\beta,\beta^*$ parameters. 

The matrix-valued $\mathcal{P}^\pm_k$ of  \eqref{eq:PsiNewNew}-\eqref{eq:PsiDaggerNewNew} are orthogonal Hermitian projection operators\footnote{These operators satisfy the mathematical requirements of projection operators.} onto the $\pm$ eigenspaces of another Hermitian operator $Q_k$, defined as:
\begin{equation} \label{eqn:Qk}
    Q_k=\gamma^5_E\big(i\gamma^\mu_E b_{k\mu}+m\bone_{D}\big)\;,
\end{equation}
with $m$ a network hyperparameter.\footnote{The operator $Q_k$ is chosen such that the Fourier transformation of its inverse is the Dirac propagator. The eigenspectrum of $Q_k$ consists of nonzero positive and negative eigenvalues, rendering $Q_k^{-1/2}$ well-defined but non-Hermitian. Therefore, a definition of the Dirac NN-FT using $Q^{-1/2}_k$ directly, as is the most straightforward approach, would not satisfy the much-necessary conjugation relation $\Psi^\dagger_i=(\Psi^\top_i)^*$. The splitting of spectral modes into positive and negative is required in order to recover the correct physics.} The hermiticity of $Q_k$ is a natural consequence of the hermiticity of $\gamma^\mu_E$ and $\gamma^5_E$ and their mutual anticommutativity. Furthermore, observe that $Q_k^2=E_k^2\,\bone_{D}$ leads to a split of the eigenvalues of $Q_k$ into $\pm E_k:=\pm \sqrt{\mathbf{b}_k^2+m^2}$, corresponding to respective domains of $\mathcal{P}^\pm_k$, which have the explicit form:
\begin{equation}
    \mathcal{P}^\pm_k=\frac{1}{2}\Big(\bone_{D}\pm\frac{Q_k}{E_k}\Big)\;.\label{eq:Projs}
\end{equation}
In physics language, $\mathcal{P}^+_k$ projects out the positive frequency eigenmodes (corresponding to ``particles''), while $\mathcal{P}^-_k$ projects out the negative frequency eigenmodes (``antiparticles''). We obtain alternate expressions for $Q_k$ and its inverse from \eqref{eq:Projs},
\begin{equation}
    Q_k=E_k(\mathcal{P}^+_k-\mathcal{P}^-_k)\;,\qquad Q_k^{-1}=\frac{1}{E_k}(\mathcal{P}^+_k-\mathcal{P}^-_k)\;\label{eq:QfromP},
\end{equation}
which will be used to compute correlation functions of $\Psi, \Psi^\dagger$ hereafter.

Next, observe that the $2$-pt function of the free Dirac NN-FT is naturally defined as the covariance between  $\Psi_E$ and its Hermitian conjugate $\Psi_E^\dagger$,\footnote{Unlike real random variables, the $2$nd moment for a complex random variable $z$ is defined as $\mathbb{E}[z\bar{z}]$.}
\begin{align}
 G^{(2)}_{\text{Dirac}}(x,y) := \mathbb{E}[\Psi_\text{E}(x)\Psi_\text{E}^\dagger(y)] =   
\left[
\begin{array}{ccc}
\mathbb{E}[\Psi_1(x)\Psi^\dagger_1(y)] & \hdots &  \mathbb{E}[\Psi_1(x)\Psi^\dagger_D(y)]  \\
\vdots & & \vdots \\
\mathbb{E}[\Psi_D(x)\Psi^\dagger_1(y)]  & \hdots & \mathbb{E}[\Psi_D(x)\Psi^\dagger_D(y)] 
\end{array}
\right].
\end{align}
The matrix elements satisfy 
\begin{align}
    \mathbb{E}[\alpha_{j_1 k_1}\alpha^*_{j_2 k_2}] &= \mathbb{E}[\beta_{j_1 k_1}\beta^*_{j_2 k_2}]= - \mathbb{E}[\alpha^*_{j_1 k_1}\alpha_{j_2 k_2}]=- \mathbb{E}[\beta^*_{j_1 k_1}\beta_{j_2 k_2}]=\sigma^2\delta_{j_1 j_2}\delta_{k_1 k_2}\;, \\
    \mathbb{E}[\alpha_{j_1 k_1}\alpha_{j_2k_2}] &=\mathbb{E}[\beta_{j_1 k_1}\beta_{j_2k_2}]=\mathbb{E}[\alpha^*_{j_1k_1}\alpha^*_{j_2k_2}]=\mathbb{E}[\beta^*_{j_1k_1}\beta^*_{j_2k_2}]=0\;,
\end{align} 
to reproduce the Euclidean free Dirac propagator of \eqref{lagrangian}. For example, the $(i,j)$-th term evaluates to
\begin{align}
    \mathbb{E}[\Psi_i(x)\Psi^\dagger_j(y)]&=\frac{\text{Vol}(B^4_\Lambda)}{N\sigma^2(2\pi)^4}\notag\times\\&~\sum_{j_1, j_2=1}^D\sum_{k_1,k_2=1}^N\Big( \mathbb{E}[\alpha_{j_1k_1}\alpha^*_{j_2k_2}]\mathbb{E}\bigg[\frac{1}{\sqrt{E_{k_1}E_{k_2}}}(\mathcal{P}^+_{k_1})_{ij_1}(\mathcal{P}^+_{k_2})_{j_2j}e^{i(b_{k_1\mu}x_\mu-b_{k_2\mu}y_\mu+c_{k_1}-c_{k_2})}\bigg]\notag
    \\&\hspace{22.5mm}+\mathbb{E}[\beta^*_{j_1k_1}\beta_{j_2k_2}]\mathbb{E}\bigg[\frac{1}{\sqrt{E_{k_1}E_{k_2}}}(\mathcal{P}^-_{k_1})_{ij_1}(\mathcal{P}^-_{k_2})_{j_2j}e^{i(b_{k_1\mu}x_\mu-b_{k_2\mu}y_\mu+c_{k_1}-c_{k_2})}\bigg]\Big)\notag\\
    &=\frac{\text{Vol}(B^4_\Lambda)}{N(2\pi)^4}\sum_{k=1}^N \mathbb{E}\bigg[\frac{1}{E_k}\big(\mathcal{P}^+_k-\mathcal{P}^-_k\big)_{ij}e^{ib_{k\mu}(x_\mu-y_\mu)}\bigg]\notag\\
    &=\frac{1}{(2\pi)^4}\int d^4 b\; \frac{e^{ib_{\mu}r_\mu}}{\big[\gamma^5_E(i\gamma^\mu_E b_{\mu}+m\bone_D)\big]_{ij}}\;.
\end{align}
The last line leverages the i.i.d.\  nature of the $b$ parameters across the $k$ index and equation~\eqref{eq:QfromP}, and introduces a shorthand notation $r_\mu:=x_\mu-y_\mu$. We also use the familiar shorthand $1/A$ to denote the inverse of a matrix $A$.

Computing the Fourier transform of the full propagator $G^{(2)}_{\text{Dirac}}(x,y)$ in terms of $p_\mu$, the momentum conjugate to $r_\mu$, gives:
\begin{align}
    G^{(2)}_{\text{Dirac}}(p)&=\int d^4b_k\, \frac{\delta^{(4)}(b_{k\mu}-p_\mu)}{\gamma^5_E(i\gamma^\mu_E b_{k\mu}+m)}=\frac{1}{\gamma^5_E(i\slashed{p}+m)}\label{propagator}\;,
\end{align}
where we define $\slashed{p}:=\gamma^\mu_E p_\mu$ as the usual Feynman slash notation in Euclidean signature. Comparison with \eqref{lagrangian} confirms that \eqref{propagator} is indeed the correct Euclidean propagator for the free Dirac theory. In the infinite width limit, the Central Limit Theorem applies, and this is the only non-trivial connected correlator, as expected for the free Dirac theory. 

Finally, although we have already obtained the propagator for the free Dirac field, we would like to reconsider how correlators behave under spatial rotations and spinor index rotations from the point of view of the NN-FT. Consider for a moment a theory of scalars, for example the architecture of the complex scalar $\phi$ in \eqref{eq:phi}. Rotating the spatial input by $R^{-1}\in SO(4)$ can be reinterpreted as a rotation on the input weights $b_{k}$:
\begin{equation}
    \phi(R^{-1}x;\,a,b,c)=\phi(x;\,a, Rb, c)\;,
\end{equation}
which also depends upon $E_\phi(b_k):=\sqrt{b_k^2+m^2}$ being invariant under rotations of $b_k$. Then, because this rotation has unit determinant and leaves $P(b)$ invariant, it follows immediately that
\begin{equation}
    \big\langle\phi(R^{-1}x)\big\rangle=\big\langle \phi(x)\big\rangle\;,
\end{equation}
demonstrating that the complex scalar has a symmetry of spatial rotations. Indeed, all correlators of $\phi$ are rotationally-invariant, as is easily shown.

On the other hand, the Dirac NN-FT contains projection operators $\mathcal{P}^\pm(b_k)$ that transform nontrivially under rotations of $b_k$. Invariance in this theory requires a \emph{simultaneous} rotation of input and output weights:
\begin{align}
    b_k&\to Rb_k\,,\notag\\
    \alpha_k&\to S(R)\alpha_k\;;\qquad\qquad \alpha^*_k\to S(R)^* \alpha^*_k\,,\label{eq:param_rotations}\\
    \beta^*_k&\to S(R)\beta^*_k\;;\qquad\qquad \beta_k\to S(R)^*\beta_k\notag\;,
\end{align}
with $S(R)$ the unitary spinor representation of $R$.\footnote{$S(R)$ is unitary, not orthogonal, so the term ``rotation'' applied to the output weights is a slight misuse of terminology.} Then, using $S(R)^{-1}\gamma^\mu_E S(R)={R^\mu}_\nu \gamma^\nu_E$ and $S(R)^{-1}\gamma^5_E S(R)=\gamma^5_E$, we obtain
\begin{equation}
    S(R) \mathcal{P}^\pm(b_k) S(R)^{-1}=\mathcal{P}^\pm(Rb_k)\;,
\end{equation}
with $E(b_k)$ still unchanged. This yields the key identities:
\begin{align}
    &S(R)\Psi(R^{-1}x;\,\alpha,\beta^*,b,c)=\Psi(x;\,S(R)\alpha, S(R)\beta^*, Rb, c)\notag\\[10 pt]
    &\Psi^\dagger(R^{-1}x;\,\alpha^*,\beta,b,c)S(R)^{-1}=\Psi^\dagger(x;\,S(R)^*\alpha^*, S(R)^*\beta, Rb, c)\;.
\end{align}
By a change of variables in parameter space with unit determinant and the invariance of $P(b)$, $P(\alpha,\alpha^*)$, and $P(\beta,\beta^*)$ under equation \eqref{eq:param_rotations}, we find
\begin{equation}
    \big\langle\Psi(x)\big\rangle=\big\langle S(R)\Psi(R^{-1}x)\big\rangle\,,\qquad\big\langle\Psi^\dagger(x)\big\rangle=\big\langle \Psi(R^{-1}x)S(R)^{-1}\big\rangle\;.
\end{equation}
More generally, for any correlator with $n$ $\Psi$'s and $m$ $\Psi^\dagger$'s,
\begin{align}
    \big\langle\Psi(x_1)\dots&\Psi(x_n)\Psi^\dagger(y_1)\dots\Psi^\dagger(y_m)\big\rangle\notag\\
    &=\Big(\prod_{a=1}^n S(R)\Big)\big\langle\Psi(R^{-1}x_1)\dots\Psi(R^{-1}x_n)\Psi^\dagger(R^{-1}y_1)\dots\Psi^\dagger(R^{-1}y_m)\big\rangle\Big(\prod_{b=1}^m S(R)^{-1}\Big)\;\label{eq:psi_invariance},
\end{align}
with each $S(R)$ acting on the appropriate spinor indices. Equation \eqref{eq:psi_invariance} demonstrates covariance under
\begin{equation}
    \Psi(x)\to \Psi'(x)=S(R)\Psi(R^{-1}x)\;\label{eq:Psiprime},
\end{equation}
which is indeed the correct symmetry transformation of Dirac spinors under rotations (c.f.~\cite{PeskinSchroeder}).

As a specific example, consider the two-point function of $\Psi$ evaluated at points $x$ and $y$ rotated via $R^{-1}$:
\begin{align}
    G^{(2)}_\Psi(R^{-1}x, R^{-1}y)&=\mathbb{E}[\Psi(R^{-1}x)\Psi^\dagger(R^{-1}y)]\notag\\[5 pt]
    &=\frac{1}{(2\pi)^4}\int d^4 b\;e^{ib_\mu (R^{-1}r)_\mu}\;\frac{1}{E(b)}\big(\mathcal{P}^+(b)-\mathcal{P}^-(b)\big)\notag\\[5 pt]
    &=\frac{1}{(2\pi)^4}\int d^4b'\; e^{ib'_\mu r_\mu}\frac{1}{E(R^{-1}b')}\big(\mathcal{P}^+(R^{-1}b')-\mathcal{P}^-(R^{-1}b')\big)\notag\\[5 pt]
    &=S(R)^{-1}\frac{1}{(2\pi)^4}\int d^4b'\; e^{ib'_\mu r_\mu}\,\frac{1}{E(b')}\big(\mathcal{P}^+(b')-\mathcal{P}^-(b')\big)\;S(R)\notag\\[5 pt]
    &=S(R)^{-1}G^{(2)}_\Psi(x,y)S(R)\;\label{eq:incorrect_rotation},
\end{align}
where the third line uses that the transformation $b\to b'=Rb$ has unit determinant. From \eqref{eq:incorrect_rotation}, it is clear that the correlator is not invariant under spatial rotations $x\to R^{-1}x$ alone (equivalently, $b\to Rb$), as is the case in a theory of scalars. Rather, we must also rotate the spinor indices of $\Psi$ via $S(R)$ to form $\Psi'(x)$, as in equation \eqref{eq:Psiprime}. The two-point function of this appropriately transformed $\Psi'$ is given by:
\begin{align}
    G^{(2)}_{\Psi'}(x,y)&=\mathbb{E}[\Psi'(x)\Psi'\,^\dagger(y)]\notag\\
    &=S(R) \mathbb{E}[\Psi(R^{-1}x)\Psi^\dagger(R^{-1}y)]S(R)^{-1}\notag\\
    &=S(R)G^{(2)}_{\Psi}(R^{-1}x, R^{-1}y)S(R)^{-1}=G^{(2)}_\Psi(x,y)\;,
\end{align}
with the last line using \eqref{eq:incorrect_rotation}.

Next, let us briefly discuss how this architecture may be extended to build interacting non-supersymmetric NN-FTs; for example, the classical Yukawa theory.

\subsection{Yukawa Couplings}
As an instance of an interacting NN-FT involving the free Dirac architecture, we engineer an architecture with an action $S_{\text{Yukawa}}$ equivalent to the classical Yukawa theory by using the principles of Section 4 from \citep{demirtas2023neural}. This architecture has hybrid outputs; i.e., it consists of complex scalar- and Grassmann-valued fields. To that end, we first append the free Dirac architecture to a complex-valued extension of the free real scalar NN-FT architecture from \citep{halverson2021building,demirtas2023neural}, followed by an induction of statistical correlations via deformations of the architecture's parameters, which leads to the deformations of its free action in the path integral. 

We begin with an asymptotically wide single hidden layer fully-connected feedforward architecture having an exponential activation function, two hyperparameters $m_b, m_f$ which play the roles of scalar and Dirac masses respectively, and i.i.d.\  parameter distributions $P_G(a, b,c, \alpha, \alpha^*, \beta, \beta^*) = P_G(a)P_G(b)P_G(c)P_G(\alpha, \alpha^*)P_G(\beta,\beta^*)$. The parameters $a$ and $\alpha,\beta$ are complex c-valued and Grassmann-valued random variables, respectively, while $b$ and $c$ are real random variables; the subscript `$G$' stands for Neural Network Gaussian Process. The densities are:
\begin{align}
P_G(\alpha, \alpha^*) &= \prod_{j=1}^{N}P_G(\bm{\alpha}_j, \bm{\alpha}^*_j), \quad \,\,  \quad P_G(\bm{\alpha}_j, \bm{\alpha}^*_j) =\sigma^{2D} \exp\bigg[{ \frac{1}{\sigma^2} \sum_{i=1}^D \alpha_{ij}\alpha^*_{ij} }\bigg], \\
P_G(\beta, \beta^*) &= \prod_{j=1}^{N}P_G(\bm{\beta}_j, \bm{\beta}^*_j), \quad \,\,  \quad P_G(\bm{\beta}_j, \bm{\beta}^*_j) =\sigma^{2D} \exp\bigg[{ \frac{1}{\sigma^2} \sum_{i=1}^D \beta_{ij}\beta^*_{ij} }\bigg], \\
P_G(a) &=\prod_{j=1}^{N}P_G(a_j),\quad ~ \quad \,\,  \quad P_G(a_j) = \frac{1}{2\pi\sigma_a^2}\exp\bigg[{ - \frac{1}{\sigma_{a}^2}{ a_j \bar{a}_j}  }\bigg], \\
P_G(b) &= \prod_{j=1}^{N} P_G(\bm{b}_j), \quad \, \quad \,\, \quad P_G(\bm{b}_j) = \text{Unif}(B^4_{\Lambda}) \quad \quad \bm{b}_j = (b_{j1},\cdots, b_{j4}), \\
    P_G(c) &= \prod_{j=1}^N P(c_j),\quad \quad \,\,\,\, \quad P(c_j) = \text{Unif}([-\pi, \pi]).
\end{align}
This NN-FT may be expressed in terms of the following shorthand notation,
\begin{align}
    y_{\alpha,\beta,a,b,c}(x) = \begin{Bmatrix}
        \Psi_{\alpha, \beta,  b, c}(x)  \\
        \phi_{a,b,c}(x)
    \end{Bmatrix} .
\end{align}
The first D components $\Psi_{ {\alpha,\beta,  b, c}}(x)$ are the Dirac spinor defined in \eqref{eq:PsiNewNew}, while the complex scalar NN field $\phi_{a,b,c}(x)$ is a trivial extension of the real scalar NN-FT in \citep{halverson2021building, demirtas2023neural}:
\begin{align}
    \phi_{a,b,c}(x) &= \sqrt{ \frac{\text{Vol}(B^4_{\Lambda}) }{N \sigma^2_a (2\pi)^4 }  } \sum_{j=1}^{N} \frac{a_{j} \exp \big[i\sum_{\mu=1}^4(b_{j\mu}x^{\mu} + c_j)\big]}{\sqrt{{\bf b}_j^2 + m_b^2}}\;,\label{eq:phi} 
\end{align}
leading to the following free action: 
\begin{align}
    S_{\text{free}}[\phi, \Psi] = \int d^4x\, \big[ - \partial^\mu {\phi}^\dagger \partial_\mu\phi + m_b^2 {\phi}^\dagger\phi + \Psi^\dagger \gamma^5_E(\gamma^E_\mu \partial_\mu + m_f)\Psi \big],
\end{align}
where we introduced shorthand notations $\Psi:=\Psi_E(x),~\Psi^\dagger:=\Psi^\dagger_E(x),~\phi:=\phi(x)$.

The next and final step requires a projection of fermionic output components as $\Psi_{L,R} = (1\pm \gamma^5_E)\Psi/2$, followed by deformations of the original i.i.d.  parameter distributions into 
\begin{align}
    P(a,b,c, \alpha, \alpha^*,\beta,\beta^*) &= P_G(a)P_G(b)P_G(c)P_G(\alpha, \alpha^*)P_G(\beta,\beta^*)\times\nonumber\\&\hspace{20mm}e^{ ~g \int d^4x \, \big( \Psi^{\dagger}_{L,\,\alpha^*,b,c}\phi_{a,b,c}\Psi_{R,\,\alpha,b,c} + \Psi^{\dagger}_{R,\,\alpha^*,b,c}\phi^\dagger_{\bar{a},b,c}\Psi_{L,\,\alpha,b,c} \big) },
\end{align}
while preserving all other architecture details and introducing another real hyperparameter~$g$. Then, the resulting action, which we call $S_{\text{Yukawa}}$, takes the following shape:
\begin{align}
    S_{\text{Yukawa}}[\phi, \Psi] =& \int d^4x\, \bigg[ - \partial^\mu {\phi}^\dagger \partial_\mu\phi + m_b^2 {\phi}^\dagger\phi + \Psi^\dagger \gamma^5_E(\gamma^E_\mu \partial_\mu + m_f)\Psi \nonumber \\
    &\hspace{14mm}- g \int d^4x \, \big( \Psi^\dagger_L\phi\Psi_R + \Psi^\dagger_R\phi^\dagger\Psi_L \big)  \bigg].
\end{align}
In this step, statistical correlations of the parameter distributions show up as perturbations of the NN-FT action through operator insertions in the path integral formulation, 
\begin{align}
    Z[J] &= \int da \, db \, dc\, d\alpha \, d\alpha^* \, d\beta\, d\beta^*\;P(a,b,c, \alpha, \alpha^*,\beta,\beta^*)\times\notag\\&\hspace{5cm} e^{\int d^4x \, \big[J_1^\dagger(x)\Psi_{\alpha,\beta, b, c}(x) + J_1\Psi^\dagger_{\alpha,\beta, b, c} + J^\dagger_2 \phi_{a,b,c} + J_2 \phi^\dagger_{a,b,c}\big] }\;. 
\end{align}
Here, $J_1:=J_1(x)$ and $J_2:=J_2(x)$ are complex anticommuting and commuting sources, respectively. This defines the classical Yukawa NN-FT architecture.

\subsection{Leading Order Dirac Spinor Interactions at Finite \textit{N}}
It was shown in Section \ref{sec:CLT} that a Grassmann-valued NNGP admits interactions at finite width $N$ that scale as $\mathcal{O}(N^{1-r/2})$ in the $r$-th connected correlator. In this section, we compute the fourth connected correlator (the first non-trivial cumulant at finite width) for the Dirac spinor architecture introduced in Section \ref{sec:dirac_spinors}.

We begin with the four-point correlator
\begin{align*}
G^{(4)}_{i_1\dots i_4}(x_1,\dots,x_4)=\langle\Psi_{i_1}(x_1)\Psi^\dagger_{i_2}(x_2)\Psi_{i_3}(x_3)\Psi^\dagger_{i_4}(x_4)\rangle\;, 
\end{align*}
where expectation is taken with respect to the network parameters $\alpha,\alpha^*,\beta,\beta^*,b,c$. This correlator is given explicitly by:
\begin{align}
    K^4\bigg\langle\sum_{k_1,\dots,k_4=1}^N\,\sum_{j_1,\dots,j_4=1}^D &\frac{e^{i(b_{k_1\mu}{x_1}_\mu+c_{k_1}+b_{k_3\mu}{x_3}_\mu+c_{k_3}-b_{k_2\mu}{x_2}_\mu - c_{k_2}-b_{k_4\mu}{x_4}_\mu - c_{k_4})}}{\sqrt{E_{k_1}\dots E_{k_4}}}\label{eq:fourthMomentInit}\\
    &\times\Big((\mathcal{P}^+_{k_1})_{i_1 j_1}\alpha_{j_1 k_1}-(\mathcal{P}^-_{k_1})_{i_1 j_1}\beta^*_{j_1k_1}\Big)\Big((\mathcal{P}^+_{k_2})_{j_2 i_2}\alpha^*_{j_2 k_2}-(\mathcal{P}^-_{k_2})_{j_2 i_2}\beta_{j_2 k_2}\Big)\notag\\
    &\times\Big((\mathcal{P}^+_{k_3})_{i_3 j_3}\alpha_{j_3 k_3}-(\mathcal{P}^-_{k_3})_{i_3 j_3}\beta^*_{j_3 k_3}\Big)\Big((\mathcal{P}^+_{k_4})_{j_4 i_4}\alpha^*_{j_4 k_4}-(\mathcal{P}^-_{k_4})_{j_4 i_4}\beta_{j_4 k_4}\Big)\bigg\rangle\notag\;,
\end{align}
with $K=\sqrt{\text{Vol}(B^4_\Lambda)/N\sigma^2(2\pi)^4}$. We can exploit Gaussianity of $P(\alpha,\alpha^*)$ to write:
\begin{align}
    \langle \alpha_{j_1 k_1}\alpha^*_{j_2 k_2}\alpha_{j_3 k_3}\alpha^*_{j_4 k_4}\rangle&=\langle \alpha_{j_1 k_1}\alpha^*_{j_2 k_2}\rangle\langle\alpha_{j_3 k_3}\alpha^*_{j_4 k_4}\rangle-\langle\alpha_{j_1 k_1}\alpha_{j_3 k_3}\rangle\langle\alpha^*_{j_2 k_2}\alpha^*_{j_4 k_4}\rangle\notag\\&\hspace{46mm}-\langle\alpha_{j_1 k_1}\alpha^*_{j_4 k_4}\rangle\langle\alpha_{j_3 k_3}\alpha^*_{j_2 k_2}\rangle\label{eq:fermionicWick}\\[5 pt]
    &=\sigma^4(\delta_{j_1 j_2}\delta_{k_1 k_2}\delta_{j_3 j_4}\delta_{k_3 k_4}-\delta_{j_1 j_4}\delta_{k_1 k_4}\delta_{j_3 j_2}\delta_{k_3 k_2})\;,\notag
\end{align}
where minus signs arise from exchanging the Grassmann variables $\alpha$. Similarly,
\begin{align}
    \langle \beta_{j_1 k_1}\beta^*_{j_2 k_2} \beta_{j_3 k_3} \beta^*_{j_4 k_4}\rangle&=\sigma^4(\delta_{j_1 j_2}\delta_{k_1 k_2}\delta_{j_3 j_4}\delta_{k_3 k_4}-\delta_{j_1 j_4}\delta_{k_1 k_4}\delta_{j_3 j_2}\delta_{k_3 k_2})\\
    \langle \alpha_{j_1 k_1} \alpha^*_{j_2 k_2} \beta_{j_3 k_3} \beta^*_{j_4 k_4}\rangle &=\sigma^4 \delta_{j_1 j_2}\delta_{k_1 k_2} \delta_{j_3 j_4} \delta_{k_3 k_4}\label{eq:fermionicWick3}\;.
\end{align}
All other fourth-order moments of $\alpha,\alpha^*,\beta,\beta^*$ vanish due to the assumptions that $\alpha$ and $\beta$ are independent and have mean zero, so the only terms that survive in \eqref{eq:fourthMomentInit} are those with every Grassmann weight $\theta$ matched to its corresponding $\theta^*$. Furthermore, using equations \eqref{eq:fermionicWick}-\eqref{eq:fermionicWick3} to sum over all $k_1,\dots,k_4$ and $j_1,\dots,j_4$ in equation \eqref{eq:fourthMomentInit} results in two leftover sums from $1$ to $N$ which we relabel $k$ and $l$. Thus, after carrying out all the contractions and simplifying, we are left with
\footnotesize
\begin{align}
    G^{(4)}_{i_1\dots i_4}=K^4\sigma^4\sum_{k,l=1}^N\Big\langle\frac{1}{E_k E_l}&\Big[(\mathcal{P}^+_k)_{i_1 i_2}(\mathcal{P}^+_l)_{i_3 i_4}e^{ib_{k\mu}x_{12\mu}}e^{ib_{l\nu}x_{34\nu}}-(\mathcal{P}^+_k)_{i_1 i_4}(\mathcal{P}^+_l)_{i_3 i_2}e^{ib_{k\mu}x_{14\mu}}e^{ib_{l\nu}x_{32\nu}}\notag\\
    &\!\!\!-(\mathcal{P}^+_k)_{i_1 i_2}(\mathcal{P}^-_l)_{i_3 i_4}e^{ib_{k\mu}x_{12\mu}}e^{ib_{l\nu}x_{34\nu}}+(\mathcal{P}^+_k)_{i_1 i_4}(\mathcal{P}^-_l)_{i_3 i_2}e^{ib_{k\mu}x_{14\mu}}e^{ib_{l\nu}x_{32\nu}}\notag\\[10 pt]
    &\!\!\!-(\mathcal{P}^-_k)_{i_1 i_2}(\mathcal{P}^+_l)_{i_3 i_4}e^{ib_{k\mu}x_{12\mu}}e^{ib_{l\nu}x_{34\nu}}+(\mathcal{P}^-_k)_{i_1 i_4}(\mathcal{P}^+_l)_{i_3 i_2}e^{ib_{k\mu}x_{14\mu}}e^{ib_{l\nu}x_{32\nu}}\notag\\[10 pt]
    &\!\!\!+(\mathcal{P}^-_k)_{i_1 i_2}(\mathcal{P}^-_l)_{i_3 i_4}e^{ib_{k\mu}x_{12\mu}}e^{ib_{l\nu}x_{34\nu}}-(\mathcal{P}^-_k)_{i_1 i_4}(\mathcal{P}^-_l)_{i_3 i_2}e^{ib_{k\mu}x_{14\mu}}e^{ib_{l\nu}x_{32\nu}}\Big]\Big\rangle\;,\label{eq:fourthMomentIntermed}
\end{align}
\normalsize
where, e.g., ${x_{12}}_\mu={x_1}_\mu-{x_2}_\mu$. Upon using \eqref{eq:QfromP}, we find that \eqref{eq:fourthMomentIntermed} simplifies to
\footnotesize
\begin{equation}
    G^{(4)}_{i_1\dots i_4}(x_1,\dots,x_4)=K^4\sigma^4\sum_{k, l}\Big\langle (Q_k^{-1})_{i_1 i_2}\,(Q_l^{-1})_{i_3 i_4}e^{ib_{k\mu}x_{12\mu}}e^{ib_{l\nu}x_{34\nu}}-(Q_k^{-1})_{i_1 i_4}\,(Q_l^{-1})_{i_3 i_2}e^{ib_{k\mu}x_{14\mu}}e^{ib_{l\nu}x_{32\nu}}\Big\rangle\;.\label{eq:fourthMomentPsi}
\end{equation}
\normalsize
The fourth connected correlator is defined as
\begin{equation}
    \kappa^{(4)}_{i_1\dots i_4}(x_1,\dots,x_4)=G^{(4)}_{i_1\dots i_4}(x_1,\dots,x_4)-\Big[G^{(2)}_{i_1 i_2}(x_1,x_2)G^{(2)}_{i_3 i_4}(x_3,x_4)-G^{(2)}_{i_1i_4}(x_1,x_4)G^{(2)}_{i_3i_2}(x_3,x_2)\Big]\label{eq:fourthCumulantInit}\;,
\end{equation}
where the $G^{(2)}_{i_1i_3}(x_1,x_3)G^{(2)}_{i_2i_4}(x_2,x_4)$ term that would exist in the bosonic case is absent because $\langle\Psi_i(x)\Psi_j(y)\rangle=0$ for all $i,j,x,y$.\footnote{To see this, note that there is a nonzero $c_k$-dependent phase in the exponential that averages to 0.} The two-point function is
\begin{equation}
    G^{(2)}_{i_1 i_2}(x_1,x_2)=K^2\sigma^2\sum_{k=1}^N \Big\langle(Q^{-1}_k)_{i_1 i_2}e^{ib_{k\mu}x_{12\mu}}\Big\rangle\label{eq:twoPointDirac}\;,
\end{equation}
which we can then use to compute the four-point connected correlator. For ease of notation, let $A_k:=(Q_k)^{-1}_{i_1 i_2}\,e^{ib_{k\mu}{x_{12}}_\mu}$, $B_k:=(Q_k)^{-1}_{i_3 i_4}\,e^{ib_{k\mu}{x_{34}}_\mu}$, $C_k:=(Q_k)^{-1}_{i_1 i_4}\,e^{ib_{k\mu}{x_{14}}_\mu}$, $D_k:=(Q_k)^{-1}_{i_3 i_2}\,e^{ib_{k\mu}{x_{32}}_\mu}$. Then
\begin{align}
    \kappa^{(4)}_{i_1\dots i_4}(x_1,\dots, x_4)&=K^4\sigma^4 \sum_{k,l=1}^N\bigg[\Big(\big\langle A_k B_l\big\rangle-\big\langle C_k D_l\big\rangle\Big)-\Big(\big\langle A_k\big\rangle\big\langle B_l\big\rangle-\big\langle C_k\big\rangle\big\langle D_l\big\rangle\Big)\bigg]\notag\\[5 pt]
    &=K^4\sigma^4 \sum_{k=1}^N\bigg[\Big(\big\langle A_k B_k\big\rangle-\big\langle C_k D_k\big\rangle\Big)-\Big(\big\langle A_k\big\rangle\big\langle B_k\big\rangle-\big\langle C_k\big\rangle\big\langle D_k\big\rangle\Big)\bigg]\notag\\[5 pt]
    &=\frac{1}{N}\frac{V^2}{(2\pi)^8}\bigg[\Big(\big\langle e^{ib_\mu({x_{12}}_\mu+{x_{34}}_\mu)}(Q^{-1})_{i_1 i_2}\,(Q^{-1})_{i_3 i_4}\,\big\rangle\notag\\&\hspace{20mm}-\big\langle e^{i b_\mu {x_{12}}_\mu} (Q^{-1})_{i_1 i_2}\big\rangle \big\langle e^{i b_\mu {x_{34}}_\mu} (Q^{-1})_{i_3 i_4}\big\rangle\Big)-\Big(2\leftrightarrow4\Big)\bigg]\;,
\end{align}
where $V:=\text{Vol}(B^4_\Lambda)$. We emphasize that the fourth connected correlator of $\Psi$ is shown to have the expected $1/N$ scaling predicted by the CLT.

\section{Supersymmetry \label{sec: SUSY}}

We now construct supersymmetric theories in one and four dimensions using a superspace formalism and concrete input layers. We will demonstrate that the results may be used to construct a large class of interacting supersymmetric theories by appending additional layers whose parameters are independent of the parameters of the SUSY input layer.

\subsection{Supersymmetric Neural Network Quantum Mechanics}
\label{sec:1dSUSY}
We now turn to the realization of SUSY at input in the simplest case: SUSY quantum mechanics\footnote{By QM we mean that we have a SUSY 1d Euclidean theory. However, in general the theory is not reflection positive (the Euclidean analog of unitarity), and therefore it is not the Wick rotation of a unitary real-time QM theory. However, RP NN-FT constructions do exist \cite{ferko2025qmnn}, and combining it with our SUSY construction here is an interesting direction for future work.}. We will realize interacting theories via an appropriate SUSY input layer construction, which can be broadened to much more general SUSY theories by appending \emph{any} independent layer after the SUSY layer.

Consider 1d superspace with coordinates $(\tau,\theta,\bar{\theta})$, with $\theta$ and $\bar{\theta}$ Grassmann-valued. The supersymmetry algebra is defined in terms of two fermionic operators $Q$ and $\bar{Q}$ that satisfy the following anticommutation relation:
\begin{equation}
    \{Q,\bar{Q}\}=2H\;,
\end{equation}
for an appropriate Hamiltonian $H$. We can then define a group element
\begin{equation}
    G(\tau,\theta,\bar{\theta})=e^{-(H\tau+\theta Q+\bar{\theta}\bar{Q})}\;,
\end{equation}
and the multiplication of group elements $G(0,\epsilon,\bar{\epsilon})\,G(\tau,\theta,\bar{\theta})$ induces the following superspace transformations:
\begin{align}
    \tau&\to\tau+\bar{\epsilon}\theta+\epsilon\bar{\theta}\notag\\
    \theta&\to\theta+\epsilon\label{eq:superspace_transfos}\\
    \bar{\theta}&\to\bar{\theta}+\bar{\epsilon}\notag\;.
\end{align}
These transformations are realized on superfields $F(\tau,\theta,\bar{\theta})$ by the differential operators:
\begin{align}
    Q&=\partial_\theta+\bar{\theta}\partial_\tau\notag\\
    \bar{Q}&=\partial_{\bar{\theta}}+\theta\partial_\tau\notag\;,
\end{align}
where we use the same letters $Q,\bar{Q}$ as the generators because they satisfy the same anticommutator:
\begin{equation}
    \{Q,\bar{Q}\}=2\partial_\tau:=2H\;.
\end{equation}
Additionally, we can define operators $D$ and $\bar{D}$:
\begin{align}
    D&=\partial_\theta-\bar{\theta}\partial_\tau\notag\\
    \bar{D}&=\partial_{\bar{\theta}}-\theta\partial_\tau\notag\;,
\end{align}
that are constructed to satisfy the following anticommutation relations:
\begin{align}
    \{D,\bar{D}\}&=-2H\notag\\
    \{D,Q\}&=\{D,\bar{Q}\}=\{\bar{D},Q\}=\{\bar{D},\bar{Q}\}=0\notag\;.
\end{align}
These $\bar{D}$ and $D$ operators will be used later to define chiral and antichiral supernetworks, respectively.

\medskip
Having set our 1d superspace conventions, let us construct a SUSY input layer.
Consider
\begin{equation}
    z_i=W_i\tau+\xi_i\theta+\bar{\xi}_i\bar{\theta}+b_i\;,
\end{equation}
with $i=1,\dots, N$. We assume that all parameters $W$, $b$, $\xi$, and $\bar{\xi}$ are i.i.d.\  across the $i$ index, with $b$, $\xi$, and $\bar{\xi}$ uniformly distributed: $P(b_i)=\text{Unif}([-\pi,\pi])$, $P(\xi_i)=A_\xi$, and $P(\bar{\xi}_i)=A_{\bar{\xi}}$, where $A_\xi,\,A_{\bar{\xi}}\in\mathbb{R}$.\footnote{The most general function of a single Grassmann variable $\xi$ is $P(\xi)=A+B\,\xi$. A ``uniform distribution'' on Grassmann variables is therefore simply $P(\xi)=A$. We note that the distribution has norm zero $\int d\xi \, P(\xi) = 0$. Nevertheless, it has a finite first moment, $\int d\xi \, \xi P(\xi) = A$, and can lead to non-trivial correlators. This is familiar from QFTs with normalizable zero modes, where the partition function vanishes unless the correlator contains enough fermionic fields to soak up the zero modes.} We leave $P(W_i)$ unspecified for now. We define the post-activation of $z_i$ as
\begin{equation}
    \phi_i=\cos(z_i)\;.\label{eq:input_layer}
\end{equation}
Denote the full set of parameters by $\Omega$, and let $X=(\tau,\theta,\bar{\theta})$ be a superspace coordinate that transforms under SUSY to $X^\epsilon=(\tau+\bar{\epsilon}\theta+\epsilon\bar{\theta},\; \theta+\epsilon,\;\bar{\theta}+\bar{\epsilon})$. Then, leaving the post-activation's dependence on $\Omega$ explicit, $\phi_i$ transforms under superspace transformations as:
\begin{align}
    \phi_i(X;\,\Omega)&\to \phi_i(X^\epsilon;\,\Omega)\notag\\[5 pt]
    &=\cos\Big(W_i\tau+(\xi_i+W_i\bar{\epsilon})\theta+(\bar{\xi}_i+W_i\epsilon)\bar{\theta}+(b_i+\xi_i\epsilon+\bar{\xi}_i\bar{\epsilon})\Big)\notag\\[5 pt]
    &:=\cos(W_i\tau+\xi'_i\theta+\bar{\xi}'_i\bar{\theta}+b'_i):=\phi_i(X;\,\Omega')\;,\label{eq:qm_shifts}
\end{align}
where we have defined the shifted parameters $\xi'$, $\bar{\xi}'$, and $b'$, as well as the full set $\Omega'$ that also includes the $W$ parameters. The key claim is that the correlators of $\phi$ are invariant under supersymmetry because the transformation from unprimed to primed parameters in equation \ref{eq:qm_shifts} leaves the probability density $P(\Omega)$ invariant and has unit Berezinian; see Appendix \ref{sec:berezinian} for details. With these two facts in hand, the $n$-th correlator of the SUSY-transformed $\phi$ is:
\begin{align}
    G^{(n)}_{i_1\dots i_n}(X^\epsilon_1,\dots, X^\epsilon_n)&:=\mathbb{E}_\Omega[\phi_{i_1}(X^\epsilon_1;\,\Omega)\cdots \phi_{i_n}(X^\epsilon_n;\,\Omega)]\notag\\[10 pt]
    &=\int d\Omega\,P(\Omega)\,\phi_{i_1}(X^\epsilon_1;\,\Omega)\cdots \phi_{i_n}(X^\epsilon_n;\,\Omega)\notag\\[10 pt]
    &=\int d\Omega\,P(\Omega)\,\phi_{i_1}(X_1;\,\Omega')\cdots \phi_{i_n}(X_n;\,\Omega')\notag\\[10 pt]
    &=\int d\Omega'\,P(\Omega')\,\phi_{i_1}(X_1;\,\Omega')\cdots \phi_{i_n}(X_n;\,\Omega')\notag\\[10 pt]
    &=G^{(n)}_{i_1\cdots i_n}(X_1,\dots, X_n)\;\label{eq:phi_invariance_qm},
\end{align}
demonstrating SUSY invariance. The first two lines are the definitions of the $n$-th correlator. The third uses equation \eqref{eq:qm_shifts}. The fourth line uses $d\Omega\,P(\Omega)=d\Omega'\,P(\Omega')$ (see Appendix \ref{sec:berezinian}) and the fact that $b_i$ integration over $[-\pi,\pi]$ is unchanged by the shift to $b_i'=b_i+\text{const}$: a consequence of the $2\pi$-periodicity of $\phi_i$. The final line leverages $\Omega'$ being a ``dummy'' integration variable.

As a specific example, the two-point correlator of $\phi_i$ is
\begin{align}
    G^{(2)}_{i_1 i_2}(X_1, X_2)&=\mathbb{E}_{W,\xi,\bar{\xi},b}[\cos(W_{i_1}\tau_1+\xi_{i_1}\theta_1+\bar{\xi}_{i_1}\bar{\theta}_1+b_{i_1})\;\cos(W_{i_2}\tau_2+\xi_{i_2}\theta_2+\bar{\xi}_{i_2}\bar{\theta}_2+b_{i_2})]\notag\\[10 pt]
    &=\frac{\delta_{i_1 i_2}}{2}\mathbb{E}_{W}\bigg[\int d\bar{\xi}\,d\xi\, A_\xi A_{\bar{\xi}}\,\cos(W\tau_{12}+\xi\theta_{12}+\bar{\xi}\bar{\theta}_{12})\bigg]\notag\\[10 pt]
    &=\frac{\delta_{i_1 i_2}}{2}A_\xi A_{\bar{\xi}}\,\theta_{12}\,\bar{\theta}_{12}\;\mathbb{E}_W[\cos(W\tau_{12})]\label{eq:neuron_twopoint}\;.
\end{align}
where, e.g., $\tau_{12}:=\tau_1-\tau_2$. In the last line, we used the identity
\begin{equation}
    \cos(W\tau_{12}+\xi\theta_{12}+\bar{\xi}\theta_{12})=\cos(W\tau_{12})(1+\xi\bar{\xi}\theta_{12}\bar{\theta}_{12})-\sin(W\tau_{12})(\xi\theta_{12}+\bar{\xi}\bar{\theta}_{12})\notag\;.
\end{equation}
One can explicitly check that the correlator $G^{(2)}_{i_1 i_2}(X_1, X_2)$ is invariant under SUSY by plugging in the superspace transformations \eqref{eq:superspace_transfos} and utilizing $(\theta_{12})^2=(\bar{\theta}_{12})^2=0$.

We can then define a particular supernetwork as
\begin{equation}
\Phi(\tau,\theta,\bar{\theta})=\frac{1}{\sqrt{N}}\sum_{i=1}^N \phi_i\;,
\end{equation}
that has a supersymmetric two-point correlator:
\begin{align}
    G^{(2)}_\Phi(X_1, X_2)&=\frac{1}{N}\sum_{i_1, i_2=1}^N G^{(2)}_{i_1 i_2}(X_1, X_2)\notag\\[10 pt]
    &=\frac{1}{2N}\sum_{i=1}^N \mathbb{E}_W[A_\xi A_{\bar{\xi}} \,\theta_{12} \bar{\theta}_{12} \cos(W\tau_{12})]\notag\\[5 pt]
    &=\frac{1}{2}A_\xi A_{\bar{\xi}}\,\theta_{12}\bar{\theta}_{12}\cos(W\tau_{12})\;.
\end{align}
Because $\Phi$ is mean-free\footnote{This is a consequence of $\mathbb{E}_b[\phi_i]=0$}, the two-point correlator $G^{(2)}_\Phi$ is also the two-point connected correlator $\kappa^{(2)}_\Phi$.

To study interactions, we wish to study the four-point connected correlator, which is related to the four-point correlator\footnote{The explicit dependence of higher order correlators on inputs $X_i$ is dropped for ease of notation; for example, the fourth correlator and fourth connected correlator depend on $X_1,X_2,X_3,X_4$.}:
\begin{equation}
    G^{(4)}_{i_1\dots i_4}=\mathbb{E}_{W,b,\xi,\bar{\xi}}[\cos(z_{i_1})\dots \cos(z_{i_4})]\label{eq:four-point}\;.
\end{equation}
For ease of notation, let $A_n:=W_{i_n}\tau_n+\xi_{i_n}\theta_n+\bar{\xi}_{i_n}\bar{\theta}_n$ such that $z_{i_n}=A_n+b_{i_n}$. We will calculate the expectation of \eqref{eq:four-point} with respect to $b$ first, considering the two possible nonzero cases:
\begin{itemize}
    \item \textbf{Case I}: $i_1=i_2\neq i_3=i_4$. It is a simple matter of using trigonometric identities and the fact that $\mathbb{E}_b[\cos(A+b)]=0$ for all $A$ to show that:
    \begin{equation}
        \mathbb{E}_b[\cos(A_1+b_{i_1})\dots\cos(A_4+b_{i_4})]=\frac{1}{4}\cos(A_1-A_2)\cos(A_3-A_4)\label{eq:four-pt_case1}\;.
    \end{equation}
    There are two additional terms analogous to \eqref{eq:four-pt_case1} that correspond to $i_1=i_3\neq i_2=i_4$ and $i_1=i_4\neq i_2=i_3$.

    \item \textbf{Case II}: $i_1=i_2=i_3=i_4$. Once again, we simply employ trigonometric identities to obtain
    \footnotesize
    \begin{align}
        \mathbb{E}_b[\cos(A_1+b_{i})\dots \cos(A_4+b_i)]&=\frac{1}{4}\Big(\cos(A_1-A_2)\cos(A_3-A_4)+\frac{1}{2}\cos(A_1-A_4+A_2-A_3)\Big)\notag\\
        &=\frac{1}{8}\big[\cos(A_1-A_2+A_3-A_4)\notag\\
        &\qquad\qquad+\cos(A_1-A_3+A_4-A_2)+\cos(A_1-A_4+A_2-A_3)\big]\notag\;.
    \end{align}
    \normalsize
\end{itemize}
The full four-point correlator of $\phi_i$ is therefore
\footnotesize
\begin{align}
    G^{(4)}_{i_1\dots i_4}&=\mathbb{E}_{W,\xi,\bar{\xi}}\bigg[\frac{1}{4}\Big[\delta_{i_1 i_2}\delta_{i_3 i_4}\cos(A_1-A_2)\cos(A_3-A_4)+(2\leftrightarrow3)+(2\leftrightarrow4)\Big]\notag\\[5 pt]
    &\quad-\frac{1}{8}\Delta_{i_1 i_2 i_3 i_4}\big[\cos(A_1-A_2+A_3-A_4)+\cos(A_1-A_3+A_4-A_2)+\cos(A_1-A_4+A_2-A_3)\big]\bigg]\notag\;,
\end{align}
\normalsize
where $\Delta_{i_1 i_2 i_3 i_4}:=\delta_{i_1 i_2}\delta_{i_2 i_3}\delta_{i_3 i_4}$ is only nonzero if all indices are equal (the minus sign in front of the $1/8$ term eliminates overcounting in this case). Then, the four-point correlator of $\Phi$ is:
\begin{align}
    G^{(4)}_\Phi&=\frac{1}{N^2}\sum_{i_1,\dots,i_4=1}^N G^{(4)}_{i_1 \dots i_4}\notag\\
    &=\frac{1}{N^2}\mathbb{E}_{W,\xi,\bar{\xi}}\Big[\sum_{i_1,\dots, i_4=1}^N\Big(\frac{1}{4}[\delta_{i_1 i_2}\delta_{i_3 i_4}\cos(A_1-A_2)\cos(A_3-A_4)+\text{2 perms}]\notag\\
    &\qquad\qquad\qquad\qquad\qquad\qquad-\frac{1}{8}\Delta_{i_1 i_2 i_3 i_4}[\cos(A_1-A_2+A_3-A_4)+\text{2 perms}]\Big)\Big]\;.\label{eq:four_point_correlator}\;
\end{align}
Interactions are described by connected four-point function $\kappa_{\Phi}^{(4)}$, which exactly cancels the first term in square brackets (the ``disconnected piece'') in equation \eqref{eq:four_point_correlator}:
\begin{align}
    \kappa_{\Phi}^{(4)}&=-\frac{1}{8N}\mathbb{E}_{W,\xi,\bar{\xi}}\big[\cos\big(W(\tau_{12}+\tau_{34})+\xi(\theta_{12}+\theta_{23})+\bar{\xi}(\bar{\theta}_{12}+\bar{\theta}_{34})\big)+\text{2 perms}\big]\notag\\[10 pt]
    &=-\frac{A_\xi A_{\bar{\xi}}}{8N}\,\mathbb{E}_W\big[(\theta_{12}+\theta_{34})(\bar{\theta}_{12}+\bar{\theta}_{34})\cos\big(W(\tau_{12}+\tau_{34})\big)+\text{2 perms}\big]\;.\label{eq:four_point_cumulant_Phi}
\end{align}
The four-point connected correlator therefore exhibits the $1/N$ scaling consistent with the CLT. Since \eqref{eq:four_point_cumulant_Phi} is invariant under SUSY (as is easily checked), a finite width $N$ in the supernetwork $\Phi$ induces supersymmetric interactions. 

\subsubsection{Chiral Supernetworks}
A chiral supernetwork $F$ is defined to satisfy the relation:
\begin{equation}
    \bar{D} F=0\;.
\end{equation}
It is convenient to define a chiral coordinate
\begin{equation}
    y:=\tau-\theta\bar{\theta}\;,
\end{equation}
which satisfies $\bar{D}y=0$. A simple modification of the SUSY post-activation $\phi_i$ is
\begin{equation}
    \phi^{c}_i=\cos(W_iy+\xi_i\theta+b_i)\;,
\end{equation}
where a straightforward calculation demonstrates that $\bar{D} \phi^{c}_i=0$. Any supernetwork $\Phi^{c}$ composed of these chiral post-activations is therefore a chiral supernetwork. The two-point and four-point connected correlators of the chiral supernetwork are:
\begin{align}
    \kappa^{(2)}_{\Phi^c}&=-\frac{A_\xi}{2}\,\theta_{12}\mathbb{E}_W\big[\sin(W y_{12})\big]\\[5 pt]
    \kappa^{(4)}_{\Phi^c}&=-\frac{A_\xi}{8N}\;\mathbb{E}_W\big[(\theta_{12}+\theta_{34})\sin\big(W(y_{12}+y_{34})\big)+\text{2 perms}\big]\;.
\end{align}
These cumulants provide the leading connected correlations of $\Phi^c$ up to $\mathcal{O}(1/N)$.

\subsection{Supersymmetric Neural Network Field Theory}
\label{sec:4dSUSY}

Having discussed SUSY quantum mechanics, the generalization to 4d SUSY field theory is straightforward once we have a 4d SUSY algebra. There have been many attempts in the literature to define supersymmetric field theories in Euclidean signature. The algebra used in this paper, equation \eqref{eq:SUSY_alg}, is consistent with the non-conjugate $\mathcal{N}=1$ algebra introduced by Lukierski and collaborators \cite{Lukierski1, Lukierski2, Lukierski3}, and it can also be regarded as the $\mathcal{N}=1$ subalgebra of the full Euclidean $\mathcal{N}=2$ SUSY algebra derived in \cite{McKeon_Sherry}. This Euclidean algebra was shown to follow from the Lorentzian $\mathcal{N}=2$ algebra via the Wick rotation rules outlined in \cite{VANNIEUWENHUIZEN199629}. Thus, there is consistency between the algebra presented in this section and any supersymmetric theory defined via van Nieuwenhuizen and Waldron's Wick rotations, in particular the Dirac spinor NN-FT introduced in Section \ref{sec:dirac_spinors}.\footnote{In trying to relate a minimal $\cN=2$ Euclidean theory to a minimal $\cN=1$ Lorentzian quantum theory, one naturally sets the central charges to zero. In doing so, negative norm states arise \cite{McKeon_Sherry}, but as discussed in \cite{Waldron_1998} this is not a problem, since they are removed from the physical Hilbert space by a quotient associated to reflection positivity. In such a case only the $\cN=1$ subalgebra is realized, further justifying the use of \eqref{eq:SUSY_alg}}

We therefore begin with the Euclidean SUSY algebra \cite{McKeon_Sherry}:
\begin{align}
    \{Q_a, \tilde{Q}_{\dot{a}}\}=2\sigma^m_{a\dot{a}}P^n \delta_{mn}\;\label{eq:SUSY_alg},
\end{align}
where $P^n$ is the momentum, and $\sigma^m_{a\dot{a}}:=(\sigma_j, i\bone)$ and $\bar{\sigma}^m_{a\dot{a}}:=(\sigma_j, -i\bone)$, with $\sigma_j$ ($j=1,2,3$) the Pauli matrices. These obey the following relation:
\begin{equation}
    {(\sigma^m \bar{\sigma}^n + \sigma^n \bar{\sigma}^m)_a}^b = 2\delta^{mn}{\delta_a}^b\;,
\end{equation}
which is a representation of the 4d Euclidean Clifford algebra \eqref{eq:cliff}. We emphasize once more that the spinors $Q$ and $\tilde{Q}$ in \eqref{eq:SUSY_alg} are independent, as opposed to the Lorentzian case where they are Hermitian conjugates of each other.

With the SUSY algebra, we can proceed as we did in the quantum mechanics case by defining a group element $G(x^m, \theta, \tilde{\theta})=\exp[i(-x^m P_m+\theta^a Q_a+\tilde{\theta}_{\dot{a}} \tilde{Q}^{\dot{a}})]$ then multiplying two group elements together to induce the following superspace transformations\footnote{We adopt the standard convention that undotted spinor indices are contracted as ${^a}_a$ and dotted spinor indices are contracted ${_{\dot{a}}}^{\dot{a}}$.}:
\begin{align}
    x^m&\to x^m+i\theta\sigma^m \tilde{\epsilon}-i\epsilon\sigma^m\tilde{\theta}\notag\\[5 pt]
    \theta&\to\theta+\epsilon\label{eq:4d_superspace}\\[5 pt]
    \tilde{\theta}&\to\tilde{\theta}+\tilde{\epsilon}\notag\;.
\end{align}
These transformations act on superfields $G(X):=G(x, \theta, \tilde{\theta})$ via the differential operators
\begin{align}
    Q_a&=\frac{\partial}{\partial \theta^a}-i\sigma^m_{a\dot{a}}\tilde{\theta}^{\dot{a}}\partial_m\notag\;,\\[5 pt]
    \tilde{Q}_{\dot{a}}&=-\frac{\partial}{\partial \tilde{\theta}^{\dot{a}}}+i\theta^a \sigma^m_{a \dot{a}}\partial_m\notag\;,
\end{align}
which once again obey $\{Q_a, \tilde{Q}_{\dot{a}}\}=2i\sigma^m_{a\dot{a}}\partial_m$,\footnote{Although $P_m:=-i\partial_m$ there is no minus sign inconsistency because multiplication of successive group elements induces motion in the opposite direction; cf.~\cite{WessBagger1992}.} where the vector index $m$ is contracted via the Euclidean metric $\delta_{mn}$. There are additional covariant derivatives $D$ and $\tilde{D}$ defined by:
\begin{align}
    D_a&=\frac{\partial}{\partial \theta^a}+i\sigma^m_{a\dot{a}}\tilde{\theta}^{\dot{a}}\partial_m\notag\\[5 pt]
    \tilde{D}_{\dot{a}}&=-\frac{\partial}{\partial\tilde{\theta}^{\dot{a}}}-i\theta^a \sigma^m_{a\dot{a}}\partial_m\notag\;,
\end{align}
which satisfy all of the expected anticommutation relations, including $\{D_a, \tilde{D}_{\dot{a}}\}=-2i\sigma^m_{a\dot{a}}\partial_m$.

Then, we can proceed by defining an input layer as an affine transformation on superspace, yielding the pre-activation:
\begin{equation}
    z_i=W_{im}x^m + \xi_i \theta + \tilde{\xi}_{i}\tilde{\theta}+b_i\;,
\end{equation}
and the post-activation, defined to be:
\begin{equation}
    \phi_i = \cos(z_i)\label{eq:ft_neuron},
\end{equation}
where alternatively one could choose any $2\pi$-periodic function instead of $\cos$. The post-activation is supersymmetric\footnote{As always, calling the post-activation ``supersymmetric'' means that it has supersymmetric correlators. It does not imply that $\phi$ itself is supersymmetric in any pointwise sense.} as long as the parameters $b$, $\xi$, and $\tilde{\xi}$ are drawn i.i.d.\  (over the $i$ index) from uniform distributions. As in the QM case, we choose $P(b_i)=\text{Unif}[(-\pi,\pi)]$, $P(\xi_i)=A_\xi$, and $P(\tilde{\xi}_i)=A_{\tilde{\xi}}$. We also demand that $W$ is drawn i.i.d.\ over the $i$ index but leave its density unspecified; however, we remark that $P(W_i)$ must be $SO(4)$ invariant so that correlators satisfy the rotational invariance required by supersymmetry.

Invariance of $\phi_i$ under SUSY can be seen by observing that under the superspace transformations \eqref{eq:4d_superspace}
\begin{equation}
    \phi_i(X;\,\Omega)\mapsto \phi_i(X;\,\Omega')=\cos(W_{im}x^m + {\xi'}_i \theta + {\tilde{\xi}'}_i\tilde{\theta}+{b'}_i)\;,
\end{equation}
where $\Omega$ denotes the set of all parameters and $\Omega'$ the set of all constant-shifted prime variables:
\begin{align}
    {\xi'}^a_i&=\xi^a_i-iW_{im}\varepsilon^{ab}\sigma^m_{b\dot{a}}\tilde{\epsilon}^{\dot{b}}\notag\\
    {\tilde{\xi}'}_{i\dot{a}}&=\tilde{\xi}_{i\dot{a}}-iW_{im}\epsilon^a \sigma^m_{a\dot{a}}\label{eq:ft_shifts}\\
    {b'}_i&=b_i+\xi_i\epsilon+\tilde{\xi}_{i}\tilde{\epsilon}\notag\;.
\end{align}
Here we use $\varepsilon$ as the 2d antisymmetric tensor that raises and lowers both dotted and undotted spinor indices. Invariance follows from \eqref{eq:ft_shifts} because the transformation that maps $\Omega\mapsto\Omega'$ leaves the parameter densities invariant and has unit Berezinian; see Appendix~\ref{sec:berezinian} for details. Therefore, any correlation function involving the $\phi_i$'s is also invariant: 
\begin{align}
    G^{(n)}_{i_1\dots i_n}(X^\epsilon_1,\dots, X^\epsilon_n)&:=\mathbb{E}_\Omega[\phi_{i_1}(X^\epsilon_1;\,\Omega)\cdots \phi_{i_n}(X^\epsilon_n;\,\Omega)]\notag\\[10 pt]
    &=\int d\Omega\,P(\Omega)\,\phi_{i_1}(X^\epsilon_1;\,\Omega)\cdots \phi_{i_n}(X^\epsilon_n;\,\Omega)\notag\\[10 pt]
    &=\int d\Omega\,P(\Omega)\,\phi_{i_1}(X_1;\,\Omega')\cdots \phi_{i_n}(X_n;\,\Omega')\notag\\[10 pt]
    &=\int d\Omega'\,P(\Omega')\,\phi_{i_1}(X_1;\,\Omega')\cdots \phi_{i_n}(X_n;\,\Omega')\notag\\[10 pt]
    &=G^{(n)}_{i_1\cdots i_n}(X_1,\dots, X_n)\;\label{eq:phi_invariance},
\end{align}
where, just as in the quantum mechanics case, $X^\epsilon=(x+i\theta\sigma\tilde{\epsilon}-i\epsilon\sigma\tilde{\theta},\,\theta+\epsilon,\,\tilde{\theta}+\tilde{\epsilon})$ is the SUSY-shifted superspace coordinate.

We may see the invariance directly in any correlation function. For instance, the two-point function of the post-activation is
\begin{equation}
    G^{(2)}_{i_1 i_2}(X_1, X_2)=\frac{\delta_{i_1 i_2}}{8}A_\xi A_{\tilde{\xi}}(\theta_{12})^2(\tilde{\theta}_{12})^2\mathbb{E}_W[\cos(W_m x^m_{12})]\;,
\end{equation}
with, e.g., $\theta_{12} = \theta_1-\theta_2$, $(\theta_{12})^2:={\theta_{12}}^a {\theta_{12}}_a$ (which is nonzero in general, as opposed to the 1d Grassmann-valued case). Clearly $\theta_{12}$ and $\tilde \theta_{12}$ are invariant under SUSY. The expectation does transform, but its transformation disappears in the correlator because the maximal number of Grassmann variables already appears in its coefficient.

A weakly interacting limit may be taken by summing the $\phi_i$'s into a superfield $\Phi$ as 
\begin{equation}
    \Phi(x, \theta, \tilde{\theta})=\frac{1}{\sqrt{N}}\sum_{i=1}^N \phi_i\;,
\end{equation}
where one may also include weights as coefficients to $\phi_i$ if one chooses.
This has a supersymmetric two-point function given by:
\begin{equation}
    G^{(2)}_\Phi(X_1, X_2)=\frac{1}{N}\sum_{i_1, i_2=1}^N G^{(2)}_{i_1 i_2}(X_1, X_2)=\frac{1}{8}A_\xi A_{\tilde{\xi}}(\theta_{12})^2 (\tilde{\theta}_{12})^2 \mathbb{E}_W[\cos(W_m x_{12}^m)]\;.
\end{equation}
The fourth connected correlator of $\Phi$ is
\begin{equation}
    \kappa^{(4)}_\Phi=-\frac{1}{32N}A_\xi A_{\tilde{\xi}}\, \mathbb{E}_W\big[(\theta_{12}+\theta_{34})^2\, (\tilde{\theta}_{12}+\tilde{\theta}_{34})^2\cos\big(W_m(x_{12}^m+x_{34}^m)\big) + 2\,\, \text{perms}\big]\;,
\end{equation}
which has the expected $1/N$ scaling characteristic of the CLT. Thus, as $N\to\infty$, $\Phi$ is a superfield drawn from a super-Gaussian process.

Finally, as in the SUSY QM case we define chiral supernetworks $\Phi^c$ by the condition $\tilde{D}\Phi^c=0$. Defining the chiral coordinate $y^m:=x^m+i\theta\sigma^m\tilde{\theta}$, we observe that a chiral post-activation is:
\begin{equation}
    \phi^c_i=\cos(W_{im}y^m + \xi_i \theta+b_i)\;,
\end{equation}
which has two-point function:
\begin{equation}
    \mathbb{E}_{W, \xi, b}[\phi_{i_1}^c \phi_{i_2}^c]=\frac{\delta_{i_1 i_2}}{4}A_\xi (\theta_{12})^2\mathbb{E}_W[\cos(W_m y_{12}^m)]\;.
\end{equation}
The chiral superfield $\Phi^c=\frac{1}{\sqrt{N}}\sum_i \phi_i^c$ therefore has two-point correlator:
\begin{equation}
    G^{(2)}_{\Phi^c}=\frac{1}{4}A_\xi (\theta_{12})^2\mathbb{E}_W[\cos(W_m y_{12}^m)]\;,
\end{equation}
and four-point connected correlator:
\begin{equation}
    \kappa^{(4)}_{\Phi^c}=-\frac{1}{16N}A_\xi \mathbb{E}_W\big[(\theta_{12}+\theta_{34})^2\cos\big(W_m (y_{12}^m + y_{34}^m)\big)+\text{2 perms}\big]\;,
\end{equation}
where, as usual, supersymmetric interactions are induced for finite $N$.

\subsection{Large Class of Interacting SUSY Theories}
We now show that SUSY is preserved when appending \emph{any} independent neural network architecture to one of the SUSY invariant neurons we constructed, in $d=1$ in \eqref{eq:input_layer} and in $d=4$ in \eqref{eq:ft_neuron}. This operation changes the correlators and yields a large class of theories.

Consider a neural network $F$ with input layer given by the SUSY post-activations $\phi_i$.  One could think of this by appending a network $G$ after $\phi_i$,
\begin{equation}
    F(X) = G(\phi_i(X)),
    \label{eqn:deep_SUSY}
\end{equation}
where $X$ is a superspace coordinate.
If the parameters in $\phi$ are independent of all remaining network parameters (and superspace dependence enters $F$ only through $\phi$), then the correlators of $F$ are supersymmetric. Concretely, let $\Omega_\phi$ be the parameters of $\phi$, $\Omega_{G}$ the remaining parameters with $\Omega_{G}\,\cap\,\Omega_\phi=\emptyset$, and $\Omega_F=\Omega_\phi\cup\Omega_{G}$. Under these assumptions, the $n$-th correlator of $F$ is: 
\begin{align}
    G^{(n)}_F(X_1,\dots, X_n)&=\mathbb{E}_{\Omega_F}[F(X_1)\dots F(X_n)]\notag\\
    &=\mathbb{E}_{\Omega_{G}}\big[\mathbb{E}_{\Omega_{\phi}}[F(X_1)\dots F(X_n)]\big]\notag\;,
\end{align}
with the argument of $\mathbb{E}_{\Omega_{G}}[\cdot]$ supersymmetric by a calculation identical to that given in equations \eqref{eq:phi_invariance_qm} and \eqref{eq:phi_invariance}. Invariance in this manner holds in either the quantum mechanics or 4d field theory case.

\subsection{Supersymmetry Breaking}

Let us briefly address how supersymmetry may be broken.

In field theory, one type of symmetry breaking occurs when the theory is deformed in a way such that the symmetry is no longer manifest in the associated ensemble of fields, as defined by a partition function. This is known as \emph{explicit} symmetry breaking. Alternatively, it may also be the case that the ensemble does have the symmetry, but the (potentially metastable) vacuum state of the system does not realize the symmetry. This is \emph{spontaneous} symmetry breaking. Consider the example of a Gaussian random variable $x$,
\begin{equation}
    P(x) \propto e^{-\frac{1}{2} x^2}
\end{equation}
which has an $x\leftrightarrow -x$ symmetry. The deformations of this density to 
\begin{equation}
    \qquad P(x) \propto e^{-\frac{1}{2} x^2 - g x^3 - \lambda x^4 } \qquad \qquad P(x) \propto e^{+\frac{1}{2} x^2 - \lambda x^4}
\end{equation}
explicitly breaks the symmetry in the first case and spontaneously breaks it in the second case. The spontaneous breaking arises because the ``wrong signed" quadratic term yields two modes (analogs of vacua) away from $x=0$ that are exchanged under $x\leftrightarrow -x$.

Similarly, explicit supersymmetry breaking happens when parameter densities are deformed in such a way that our superspace mechanism on input parameters is not operative. On the other hand, if the parameter density does allow for the SUSY mechanism in the ensemble, it could be that networks that locally maximize probability in function space (the analog of metastable vacua) nevertheless break supersymmetry. 

To illustrate spontaneous SUSY breaking,  consider a simple SUSY QM case 
\begin{equation}
    \phi = \cos(W \tau + \xi \theta + \bar\xi \bar \theta + b)
\end{equation}
where SUSY arises via the mechanism of Section \ref{sec:1dSUSY}; via uniform $P(b)$, $P(\xi)$, and $P(\bar\xi)$. If the conditions are violated, the ensemble violates the symmetry, SUSY is explicitly broken.

Let us consider spontaneous breaking of supersymmetry, expanding the parameters around values $W_*$, $b_*$, $\xi_*$, $\bar\xi_*$ which are either local maxima of the densities or lying on a flat direction, e.g. in the case of uniform densities. By the ``value" of the Grassmann variables, we mean the value that has been implicit throughout, as in Footnote \ref{footnote:grassmann_subtlety}.
Expanding in fluctuations around these values, e.g. as $W = W_* + \delta W$ and similarly for the other variables, we have 
\begin{equation}
    \phi = \cos\left((W_* +\delta W) \tau + (\xi_* + \delta \xi) \theta + (\bar\xi_* + \delta\bar\xi) \bar \theta + (b_*+\delta b)\right).
\end{equation}
Ignoring the fluctuation, the mode is 
\begin{equation}
    \phi = \cos(W_* \tau + \xi_* \theta + \bar\xi_* \bar \theta + b_*),
\end{equation}
which is not translation invariant in $\tau$, $\theta$, or $\bar\theta$; SUSY is spontaneously broken, and so is ordinary translation invariance in $\tau$.  This should be contrasted to the case of $P(W)$ where $W_* = 0$, in which restricting to that mode in $W$ yields
\begin{equation}
    \phi = \cos(\xi_* \theta + \bar\xi_* \bar \theta + b_*),
\end{equation}
which is invariant under $\tau$ translations, but SUSY is still spontaneously broken. Finally, if $\xi_*=\bar \xi_* = 0$ we have
\begin{equation}
    \phi = \cos(b_*)
\end{equation}
and SUSY is neither explicitly nor spontaneously broken. The former arises from the densities, whereas the later arises from the mode we expand around.

Admittedly, this formulation of SUSY breaking is presented in a way that naturally utilizes the NN-FT description, rather than via conventional means, such as the expansion into component fields and associated expectation values of associated F-terms and D-terms. We would like to have a more systematic understanding of SUSY breaking in NN-FT, including via conventional descriptions. This is as an important direction for future work.
\section{Conclusions}
\label{sec:conclusions}

In this paper we have introduced fermions into the NN-FT correspondence.

We began with the development of Grassmann-valued neural networks. First we provided a generalization of the central limit theorem to Grassmann variables in Section \ref{sec:CLT}, beginning with the discrete case and proceeding to the continuous case. In Section \ref{sec:GrassmannNNFT} we then defined two different types of Grassmann-valued NN-FTs, according to whether the output weights or the last-layer post-activations are Grassmann-valued. This amounts to the question of whether the Grassmann parameters exist ``down the network" or at the output layer. We developed some essential formalism in each of the cases.

With the foundation laid for Grassmann-valued NN-FTs, we turned to applications, including classic results and supersymmetric theories. There are three results in Section \ref{sec:applications}. First, we demonstrated an architecture that in the large-$N$ limit realized the free Euclidean Dirac spinor in four dimensions. This demonstrates the ability of NN-FT to realize classic fermionic theories. Second, we demonstrated that the Dirac spinor may be coupled to scalars via Yukawa couplings by independence breaking in the parameter densities associated to the fermionic and scalar neural networks. Third, we computed the leading four fermion interaction arising at finite-width by computing the connected $4$-point function. Via a Grassmann-valued analog of the Edgeworth expansion, one can likely compute the four-point coupling associated to this theory; see \cite{PhysRevE.104.064301,demirtas2023neural} for the analogous computation in the scalar case. 

Finally, in Section \ref{sec: SUSY} we introduced constructions of supersymmetric theories. The essential idea of the mechanism, carried out in both 1d and 4d, is to consider supersymmetry as a spacetime symmetry. Since $\{\bar Q,Q\}$ acts as a translation, it is natural to try to engineer SUSY in a manner similar to translation invariance. Translation invariance was achieved in \cite{Halverson_2021} for a wide class of interacting NN-FTs, building on the seminal idea of random Fourier features \cite{rahimi2007random} from the ML literature. Our mechanism for SUSY amounts to an extension of the input to superspace, so that the input layer is an affine transformation of superspace, and invariance properties of the associated parameter densities ensure SUSY of the input-layer post-activations. Any network with independent parameters acting on this layer gives a SUSY theory. In general these deep SUSY NN-FTs are interacting, but we also demonstrated the existence of a large-$N$ limit in which the theory is a super-Gaussian process. The network is generally a superfield, but constraints can be imposed that turn it into a chiral superfield. Superfield correlators are computed in some of the examples. We ended with a simple example exhibiting spontaneous SUSY breaking, but a more systematic understanding is left as an important direction for future work.

\vspace{1cm} \noindent \textbf{Acknowledgements.} We thank Casey Pancoast and Marc Syvaeri for collaboration during early stages of this project. This work is supported by the National Science Foundation under Cooperative Agreement PHY-2019786 (the NSF AI Institute for Artificial Intelligence and Fundamental
Interactions). J.H.\ is supported by NSF CAREER grant PHY-1848089. F.R.\ is supported by NSF grant PHY-2210333 and startup funding from Northeastern University. This research was supported in part by Perimeter Institute for Theoretical Physics. Research at Perimeter Institute is supported by the Government of Canada through the Department of Innovation, Science and Economic Development and by the Province of Ontario through the Ministry of Research, Innovation and Science. This research was supported in part by grant NSF PHY-2309135 to the Kavli Institute for Theoretical Physics (KITP).

\begin{appendices}

\section{$W[J]$ Identities for Grassmann Variables \label{app:CLT}}
Let us consider two Grassmann-valued random vectors: $\vec{X}$ and $\vec{Y}$, randomly drawn from independent distributions with densities $P(\vec{X})$ and $P(\vec{Y})$, respectively. Then, the cumulant generating function of the sum is
\begin{align}
    W_{\Vec{X} + \vec{Y}}[\vec{J}] =& \log Z_{\vec{X} + \vec{Y}}[\vec{J}] = \, \log \mathbb{E}_{P(\vec{X}, \vec{Y})}\big[e^{J_{j} X_{j }+ J_{j} Y_{j}} \big], \nonumber \\
    =& \log \mathbb{E}_{P(\vec{X})P( \vec{Y})}\big[e^{J_{j} X_{j} + J_{j} Y_{j}} \big], \nonumber \\
    = & \log \mathbb{E}_{P(\vec{X})P( \vec{Y})}\big[e^{J_{j_1} X_{j_1}}e^{ J_{j_2} Y_{j_2} } \big], \nonumber \\
    = & \log \Big[ \mathbb{E}_{P(\vec{X})}\big[e^{J_{j_1} X_{j_1}}\big] \,\mathbb{E}_{P( \vec{Y})}\big[ e^{ J_{j_2} Y_{j_2} } \big] \Big], \nonumber \\
    = & \log \mathbb{E}_{P(\vec{X})}\big[e^{J_{j_1} X_{j_1}}\big] + \log\mathbb{E}_{P( \vec{Y})}\big[ e^{ J_{j_2} Y_{j_2}} \big] , \nonumber \\
    =& W_{\Vec{X} }[\vec{J}] + W_{\vec{Y}}[\vec{J}]\;. \label{eq:sumGrassmann}
\end{align}
We used $P(\vec{X},\vec{Y}) = P(\vec{X})P( \vec{Y})$ for independent random vectors in the second line, and the Baker-Campbell-Haussdorff formula 
\begin{align}
    e^{J_{j_1} X_{j_1}} e^{X_{j_2} Y_{j_2}} = & \, e^{J_{j_1} X_{j_1} + J_{j_2} Y_{j_2} + \frac{1}{2}\big[J_{j_1} X_{j_1}, \, J_{j_2} Y_{j_2}\big] + \, \cdots } = \, e^{J_{j_1} X_{j_1} + J_{j_2} Y_{j_2} }\notag\;,
\end{align} 
as $\big[J_{j_1} X_{j_1},  J_{j_2} Y_{j_2}\big] = 0$ for $c$-numbers $J_{j_1} X_{j_1}$ and $J_{j_2} Y_{j_2}$. 

We can also deduce the cumulant generating function of the product of $\Vec{X}$ with a constant $c$-number $t$:
\begin{align}
    W_{t \Vec{X} }[\vec{J}] =& \log Z_{t \vec{X}}[\vec{J}]
    = \, \log \mathbb{E}_{P( \vec{X})}\big[e^{t ( J_j X_j ) } \big]=  W_{\Vec{X}}[t \vec{J}]\;. \label{eq:dotGrassmann}
\end{align}

Let us consider two Grassmann-valued random functions of a continuous $c$-number $x\in\mathbb{R}^d$: $\vec{X}(x)$ and $\vec{Y}(x)$. We assume that they are drawn from independent distributions $P(\vec{X}(x))$ and $P(\vec{Y}(x))$, respectively. Then, the cumulant generating functional of the sum is
\begin{align}
    &W_{\Vec{X}(x) + \vec{Y}(x)}[\vec{J}(x)] \notag \\[5 pt]
    &\qquad= \log Z_{\vec{X}(x) + \vec{Y}(x)}[\vec{J}(x)],\notag \\[5 pt]
    &\qquad = \log \mathbb{E}_{P(\vec{X}(x),\,\vec{Y}(x))}\bigg[\exp\Big(\int d^dx \big(J(x) X(x) + J(x)Y(x)\big) \Big) \bigg], \notag \\[5 pt]
    &\qquad =\log \Bigg(\mathbb{E}_{P(\vec{X}(x))}\bigg[\exp\Big(\int d^dx\, J(x) X(x)\Big)\bigg]\mathbb{E}_{P(\vec{Y}(x))}\bigg[\exp\Big(\int d^dx\, J(x) Y(x) \Big) \bigg]\Bigg), \notag \\[5 pt]
    &\qquad=\log \mathbb{E}_{P(\vec{X}(x))}\bigg[\exp\Big(\int d^dx\, J(x) X(x)\Big)\bigg]+\log\mathbb{E}_{P(\vec{Y}(x))}\bigg[\exp\Big(\int d^dx\, J(x) Y(x) \Big) \bigg], \notag \\[5 pt]
    &\qquad=W_{\vec{X}(x)}[\vec{J}(x)]+W_{\vec{Y}(x)}[\vec{J}(x)]\;. \label{eq:sumGrassmanncont}
\end{align}
We used the Baker-Campbell-Haussdorff formula in the fourth line:
\begin{align}
    e^{\int d^dx\vec{J}(x)\vec{X}(x)}\;    e^{\int d^dx\vec{J}(x)\vec{Y}(x)}&=e^{\int\big( d^dx\; \vec{J}(x)\vec{X}(x)+\vec{J}(x) \vec{Y}(x)+\frac{1}{2}\big[\vec{J}(x)\vec{X}(x),\text{ }\vec{J}(x)\vec{Y}(x)\big]+\dots\big)}\notag\\
    &=e^{\int d^dx \big(\vec{J}(x)\vec{X}(x)+\vec{J}(x)\vec{Y}(x)\big)}\notag\;,
\end{align}
which applies in this case since the product of two Grassmann variables is a $c$-number, so $[J(x)X(x),\,J(x)Y(x)]=0$ for all $x$.

Next, we deduce the cumulant generating functional of the product of $\vec{X}(x)$ with a constant $c$-number $t$:
\begin{align}
    W_{t\vec{X}(x)}[\vec{J}(x)]&=\log Z_{t\vec{X}(x)}[\vec{J}(x)]=\log\mathbb{E}_{P(\vec{X}(x))}\bigg[e^{\int d^dx\big(t(\vec{J}(x)\vec{X}(x))\big)}\bigg]=W_{\vec{X}(x)}[t\vec{J}(x)]\;. \label{eq:dotGrassmanncont}
\end{align}

\section{Proof of Unit Berezinian for SUSY NN-FT}\label{sec:berezinian}

\subsection*{Quantum Mechanics Case}
It was shown in Equation \eqref{eq:qm_shifts} that there exists a linear transformation $J$ (not to be confused with a source $J(x)$) that acts on input parameters $\Omega$ such that $\phi(X^\epsilon;\;\Omega)=\phi(X;\;\Omega')$, where $\Omega'=J\Omega$ (here we are being explicit about the dependence of $\phi$ on the network parameters). We now prove that $d\Omega'=d\Omega$.
This is a change of integration variable from $\Omega$ to $\Omega'$ induced by $J$. When working with Grassmann even and odd variables, the analog of the Jacobian is known as the Berezinian, and for transformations of the form
\begin{equation}
    J=\begin{pmatrix}
        A&B\\C&D
    \end{pmatrix}\;,
\end{equation}
the Berezinian is
\begin{equation}
    \text{Ber}(J)=\frac{\det(A-BD^{-1}C)}{\det D}\;.
\end{equation}
Our goal is therefore to show that $\text{Ber}(J)=1$. In this case, $J$ acts on vectors $(W, b, \xi, \bar{\xi})^T$ (we drop the neural index $i$ since $J$ acts independently on each neuron). We can read off each block $A$-$D$ from equation \eqref{eq:qm_shifts}:
\begin{align}
    A&=\frac{\partial(W',  b')}{\partial(W, b)}=\bone_2\notag\\[5 pt]
    B&=\frac{\partial(W',  b')}{\partial(\xi, \bar{\xi})}=\begin{pmatrix}
        0&0\\\epsilon&\bar{\epsilon}
    \end{pmatrix}\notag\\[5 pt]
    C&=\frac{\partial(\xi', \bar{\xi}')}{\partial (W,b)}=\begin{pmatrix}
        \bar{\epsilon}&0\\\epsilon&0
    \end{pmatrix}\notag\\[5 pt]
    D&=\frac{\partial(\xi', \bar{\xi}')}{\partial(\xi, \bar{\xi})}=\bone_2\;.
\end{align}
Therefore, $\text{Ber}(J)=1$ and $d\Omega'=d\Omega$. Invariance of $\phi$ correlators also requires that the parameter density is invariant; i.e., that $P(\Omega)=P(\Omega')$. This result is immediately apparent once we note that $W$ does not transform, and the distributions of $b$, $\xi$, and $\bar{\xi}$ are uniform (i.e., do not depend on $b$, $\xi$, or $\bar{\xi}$).

We emphasize that invariance of correlators requires not only invariance of the parameter density but also $\text{Ber}(J)=1$.\footnote{In general, one only needs the product $d\Omega\,P(\Omega)$ to be invariant, but in this particular case both $d\Omega$ and $P(\Omega)$ are independently invariant.}

\subsection*{Field Theory Case}
The field theory case follows the same logic as the quantum mechanics case. The Berezinian transformation that induces equation \eqref{eq:ft_shifts} has blocks $A$, $B$, $C$, and $D$ given by:
\begin{align}
    A&=\frac{\partial(W_m', b')}{\partial(W_m, b)}=\bone_5\notag\\[10 pt]
   B&=\frac{\partial(W'_m,b')}{\partial(\xi,\tilde\xi)}
=\begin{pmatrix}
  \multicolumn{4}{c}{\bzero_{4}}\\[2pt]
  \epsilon_1 & \epsilon_2 & {\tilde{\epsilon}}^{\dot 1} & {\tilde{\epsilon}}^{\dot 2}
\end{pmatrix}\notag\\[10 pt]
    C&=\frac{\partial(\xi', \tilde{\xi}')}{\partial(W_m, b)}=\begin{pmatrix}
        [-i\varepsilon^{ab}\sigma^m_{b\dot{a}}\tilde{\epsilon}^{\dot{a}}]_{2\times4}&\bzero_{2\times1}\\
        [-i\epsilon^a\sigma^m_{a\dot{a}}]_{2\times4}&\bzero_{2\times1}
    \end{pmatrix}\label{eq:Jft_blocks}\\[10 pt]
    D&=\frac{\partial(\xi', \tilde{\xi}')}{\partial(\xi, \tilde{\xi})}=\bone_4\notag\;.
\end{align}
The Berezinian is therefore:
\begin{equation}
    \text{Ber}(J)=\frac{\det(A-BD^{-1}C)}{\det D}=\det(\bone_5-BC)\;.
\end{equation}
The matrix $B$ in equation \eqref{eq:Jft_blocks} is $B=e_5\,r^T$, where $e_5$ is the fifth standard unit basis vector, and $r^T=(\epsilon_1, \epsilon_2, \tilde\epsilon^{\dot1}, \tilde\epsilon^{\dot2})$. Then $BC=e_5\,r^TC=e_5\,(r^TC):=e_5\,v^T$. An important identity is:
\begin{equation}
    v^Te_5=r^TCe_5=r^T(Ce_5)=0\label{eq:vte5}\;,
\end{equation}
where the last equation holds because the fifth column of $C$ is zero. From \eqref{eq:vte5}, we have:
\begin{equation}
    (BC)^2=(e_5v^T)(e_5v^T)=e_5(v^Te_5)v^T=0\label{eq:bc_squared}\;,
\end{equation}
as well as
\begin{equation}
    \text{tr}(BC)=\text{tr}(e_5v^T)=\text{tr}(v^Te_5)=0\label{eq:bc_traceless}\;,
\end{equation}
where the second equality follows from cyclicity of the trace. A consequence of \eqref{eq:bc_squared} is:
\begin{equation}
    \log(\bone_5-BC)=-BC\notag\;,
\end{equation}
since the power series truncates at linear order. Then, using a standard matrix identity for the determinant, we arrive at:
\begin{equation}
    \text{Ber}(J)=\det(\bone_5-BC)=\exp\big[\text{tr}\big(\log(\bone_5-BC)\big)\big]=\exp(0)=1\;,
\end{equation}
which completes the proof. Because the supersymmetric change of variables has $\text{Ber}(J)=1$ and leaves the input parameter density invariant, the correlators of any neural network driven by the supersymmetric post-activation $\phi$ are also supersymmetric.

\end{appendices}

\bibliography{refs}            

@article{Mehtafermions,
  title = {Euclidean continuation of the Dirac fermion},
  author = {Mehta, Mayank R.},
  journal = {Phys. Rev. Lett.},
  volume = {65},
  issue = {16},
  pages = {1983--1986},
  numpages = {0},
  year = {1990},
  month = {Oct},
  publisher = {American Physical Society},
  doi = {10.1103/PhysRevLett.65.1983}
}

@preprint{ringel2025applications,
  title={Applications of Statistical Field Theory in Deep Learning},
  author={Ringel, Zohar and Rubin, Noa and Mor, Edo and Helias, Moritz and Seroussi, Inbar},
  eprint={2502.18553},
  archivePrefix={arXiv},
  primaryClass={stat.ML},  
  year={2025}
}

@article{halverson2024tasi,
  title={TASI lectures on physics for machine learning},
  author={Halverson, Jim},
  eprint={2408.00082},
  archivePrefix={arXiv},
  primaryClass={hep-th},
  year={2024}
}

@book{roberts2022principles,
  title={The principles of deep learning theory},
  author={Roberts, Daniel A and Yaida, Sho and Hanin, Boris},
  volume={46},
  year={2022},
  publisher={Cambridge University Press Cambridge, MA, USA}
}

@article{PhysRevE.104.064301,
  title = {Predicting the outputs of finite deep neural networks trained with noisy gradients},
  author = {Naveh, Gadi and Ben David, Oded and Sompolinsky, Haim and Ringel, Zohar},
  journal = {Phys. Rev. E},
  volume = {104},
  issue = {6},
  pages = {064301},
  numpages = {19},
  year = {2021},
  month = {Dec},
  publisher = {American Physical Society},
  doi = {10.1103/PhysRevE.104.064301},
  url = {https://link.aps.org/doi/10.1103/PhysRevE.104.064301}
}

@inproceedings{rahimi2007random,
  title={Random features for large-scale kernel machines},
  author={Rahimi, Ali and Recht, Ben},
  booktitle={Advances in neural information processing systems},
  volume={20},
  year={2007}
}

@article{VANNIEUWENHUIZEN199629,
title = {On Euclidean spinors and Wick rotations},
journal = {Physics Letters B},
volume = {389},
number = {1},
pages = {29-36},
year = {1996},
issn = {0370-2693},
doi = {https://doi.org/10.1016/S0370-2693(96)01251-8},
author = {Peter {van Nieuwenhuizen} and Andrew Waldron}
}

@article{Waldron_1998,
   title={A Wick rotation for spinor fields: the canonical approach},
   volume={433},
   ISSN={0370-2693},
   DOI={10.1016/s0370-2693(98)00596-6},
   number={3–4},
   journal={Physics Letters B},
   publisher={Elsevier BV},
   author={Waldron, Andrew},
   year={1998},
   month=aug, pages={369–376} }

@article{Nicolai:1978vc,
    author = "Nicolai, H.",
    title = "{A Possible constructive approach to (super-$\phi^3$)$_4$: (I). Euclidean formulation of the model}",
    reportNumber = "PRINT-78-0592 (KARLSRUHE)",
    doi = "10.1016/0550-3213(78)90537-0",
    journal = "Nucl. Phys. B",
    volume = "140",
    pages = "294--300",
    year = "1978"
}

@article{ZUMINO1977369,
title = {Euclidean supersymmetry and the many-instanton problem},
journal = {Physics Letters B},
volume = {69},
number = {3},
pages = {369-371},
year = {1977},
issn = {0370-2693},
doi = {https://doi.org/10.1016/0370-2693(77)90568-8},
author = {B. Zumino}
}

@article{SchwingerFermions,
  title = {Euclidean Quantum Electrodynamics},
  author = {Schwinger, Julian},
  journal = {Phys. Rev.},
  volume = {115},
  issue = {3},
  pages = {721--731},
  numpages = {0},
  year = {1959},
  month = {Aug},
  publisher = {American Physical Society},
  doi = {10.1103/PhysRev.115.721}
}

@article{OSfermions,
  title = {Feynman-Kac Formula for Euclidean Fermi and Bose Fields},
  author = {Osterwalder, Konrad and Schrader, Robert},
  journal = {Phys. Rev. Lett.},
  volume = {29},
  issue = {20},
  pages = {1423--1425},
  numpages = {0},
  year = {1972},
  month = {Nov},
  publisher = {American Physical Society},
  doi = {10.1103/PhysRevLett.29.1423}
}

@misc{demirtas2023neural,
      title={Neural Network Field Theories: Non-Gaussianity, Actions, and Locality}, 
      author={Mehmet Demirtas and James Halverson and Anindita Maiti and Matthew D. Schwartz and Keegan Stoner},
      year={2023},
      eprint={2307.03223},
      archivePrefix={arXiv},
      primaryClass={hep-th}
}

@article{schoenholz2017correspondence,
	title        = {A Correspondence Between Random Neural Networks and Statistical Field Theory},
	author       = {Samuel S. Schoenholz and Jeffrey Pennington and Jascha Sohl-Dickstein},
	year         = 2017,
	eprint       = {1710.06570},
	archiveprefix = {arXiv},
	primaryclass = {stat.ML}
}

@article{Osterwalder1973,
	title        = {Axioms for Euclidean Green's functions},
	author       = {Osterwalder, Konrad and Schrader, Robert},
	year         = 1973,
	month        = {Jun},
	day          = {01},
	journal      = {Communications in Mathematical Physics},
	volume       = 31,
	number       = 2,
	pages        = {83--112},
	doi          = {10.1007/BF01645738},
	issn         = {1432-0916}
}

@article{maiti2021symmetryviaduality,
	title        = {Symmetry-via-Duality: Invariant Neural Network Densities from Parameter-Space Correlators},
	author       = {Anindita Maiti and Keegan Stoner and James Halverson},
	year         = 2021,
	eprint       = {2106.00694},
	archiveprefix = {arXiv},
	primaryclass = {cs.LG}
}

@article{erdmenger2021quantifying,
	title        = {Towards quantifying information flows: relative entropy in deep neural networks and the renormalization group},
	author       = {Johanna Erdmenger and Kevin T. Grosvenor and Ro Jefferson},
	year         = 2021,
	eprint       = {2107.06898},
	archiveprefix = {arXiv},
	primaryclass = {hep-th}
}

@article{Erbin:2021kqf,
    author = "Erbin, Harold and Lahoche, Vincent and Samary, Dine Ousmane",
    title = "{Non-perturbative renormalization for the neural network-QFT correspondence}",
    eprint = "2108.01403",
    archivePrefix = "arXiv",
    primaryClass = "hep-th",
    reportNumber = "MIT-CTP/5309",
    doi = "10.1088/2632-2153/ac4f69",
    journal = "Mach. Learn. Sci. Tech.",
    volume = "3",
    number = "1",
    pages = "015027",
    year = "2022"
}

@article{Howard_2025,
doi = {10.1088/2632-2153/adc8fc},
url = {https://doi.org/10.1088/2632-2153/adc8fc},
year = {2025},
month = {may},
publisher = {IOP Publishing},
volume = {6},
number = {2},
pages = {025038},
author = {Howard, Jessica N and Jefferson, Ro and Maiti, Anindita and Ringel, Zohar},
title = {Wilsonian renormalization of neural network Gaussian processes*},
journal = {Machine Learning: Science and Technology}
}

@misc{berman2022inverse,
	title        = {The Inverse of Exact Renormalization Group Flows as Statistical Inference},
	author       = {David S. Berman and Marc S. Klinger},
	year         = 2022,
	eprint       = {2212.11379},
	archiveprefix = {arXiv},
	primaryclass = {hep-th}
}

@article{Halverson_2021,
  title     = {Neural Networks and Quantum Field Theory},
  author    = {Halverson, James and Maiti, Anindita and Stoner, Keegan},
  journal   = {Machine Learning: Science and Technology},
  year      = {2021},
  volume    = {2},
  number    = {3},
  pages     = {035002},
  doi       = {10.1088/2632-2153/abeca3},
  publisher = {IOP Publishing}
}

@inproceedings{williams,
 author = {Williams, Christopher K I},
 booktitle = {Advances in Neural Information Processing Systems},
 editor = {M.C. Mozer and M. Jordan and T. Petsche},
 pages = {},
 publisher = {MIT Press},
 title = {Computing with Infinite Networks},
 url = {https://proceedings.neurips.cc/paper_files/paper/1996/file/ae5e3ce40e0404a45ecacaaf05e5f735-Paper.pdf},
 volume = {9},
 year = {1996}
}

@phdthesis{neal,
  title   = {Bayesian Learning for Neural Networks},
  author  = {Neal, Radford M.},
  school  = {Department of Computer Science, University of Toronto},
  address = {Toronto, Canada},
  year    = {1995},
  url     = {https://link.springer.com/book/10.1007/978-1-4612-0745-0}
}

@misc{yangTP1,
  title         = {Tensor Programs I: Wide Feedforward or Recurrent Neural Networks of Any Architecture are Gaussian Processes},
  author        = {Yang, Greg},
  year          = {2019},
  eprint        = {1910.12478},
  archivePrefix = {arXiv},
  primaryClass  = {cs.NE}
}

@misc{yangTP2,
  title         = {Tensor Programs II: Neural Tangent Kernel for Any Architecture},
  author        = {Yang, Greg},
  year          = {2020},
  eprint        = {2006.14548},
  archivePrefix = {arXiv},
  primaryClass  = {cs.LG}
}

@misc{Matthews2018GaussianPB,
  title         = {Gaussian Process Behaviour in Wide Deep Neural Networks},
  author        = {Matthews, Alexander G. de G. and Rowland, Mark and Hron, Jiri and Turner, Richard E. and Ghahramani, Zoubin},
  year          = {2018},
  eprint        = {1804.11271},
  archivePrefix = {arXiv},
  primaryClass  = {stat.ML}
}

@misc{Novak2018BayesianCN,
  title        = {Bayesian Convolutional Neural Networks with Many Channels are Gaussian Processes},
  author       = {Novak, Roman and Xiao, Lechao and Lee, Jaehoon and Bahri, Yasaman and Abolafia, Daniel A. and Pennington, Jeffrey and Sohl-Dickstein, Jascha},
  year         = {2018},
  eprint       = {1810.05148},
  archivePrefix= {arXiv},
  primaryClass = {cs.LG}
}

@article{GarrigaAlonso2019DeepCN,
  title        = {Deep Convolutional Networks as Shallow Gaussian Processes},
  author       = {Garriga-Alonso, Adrià and Aitchison, Laurence and Rasmussen, Carl Edward},
  year         = {2019},
  eprint       = {1808.05587},
  archivePrefix= {arXiv},
  primaryClass = {cs.LG}
}

@inproceedings{Jacot2018NeuralTK,
	title        = {Neural Tangent Kernel: Convergence and Generalization in Neural Networks},
	author       = {Arthur Jacot and Franck Gabriel and Cl{\'e}ment Hongler},
	year         = 2018,
	booktitle    = {NeurIPS}
}

@article{halverson2021building,
	title        = {Building Quantum Field Theories Out of Neurons},
	author       = {James Halverson},
	year         = 2021,
	eprint       = {2112.04527},
	archiveprefix = {arXiv},
	primaryclass = {hep-th}
}

@misc{Dyer2020AsymptoticsOW,
  title         = {Asymptotics of Wide Networks from Feynman Diagrams},
  author        = {Dyer, Ethan and Gur-Ari, Guy},
  year          = {2019},
  eprint        = {1909.11304},
  archivePrefix = {arXiv},
  primaryClass  = {cs.LG}
}

@misc{banta2023structures,
	title        = {Structures of Neural Network Effective Theories},
	author       = {Ian Banta and Tianji Cai and Nathaniel Craig and Zhengkang Zhang},
	year         = 2023,
	eprint       = {2305.02334},
	archiveprefix = {arXiv},
	primaryclass = {hep-th}
}

@article{erbin2023functional,
      title={Functional renormalization group for signal detection and stochastic ergodicity breaking}, 
      author={Harold Erbin and Riccardo Finotello and Bio Wahabou Kpera and Vincent Lahoche and Dine Ousmane Samary},
      year={2023},
      eprint={2310.07499},
      archivePrefix={arXiv},
      primaryClass={hep-th}
}

@article{Cotler:2022fze,
    author = "Cotler, Jordan and Rezchikov, Semon",
    title = "{Renormalization group flow as optimal transport}",
    eprint = "2202.11737",
    archivePrefix = "arXiv",
    primaryClass = "hep-th",
    doi = "10.1103/PhysRevD.108.025003",
    journal = "Phys. Rev. D",
    volume = "108",
    number = "2",
    pages = "025003",
    year = "2023"
}

@misc{Yaida2019NonGaussianPA,
  title         = {Non-Gaussian Processes and Neural Networks at Finite Widths},
  author        = {Yaida, Sho},
  year          = {2019},
  eprint        = {1910.00019},
  archivePrefix = {arXiv},
  primaryClass  = {stat.ML}
}

@misc{dinan2023effectivetheorytransformersinitialization,
      title={Effective Theory of Transformers at Initialization}, 
      author={Emily Dinan and Sho Yaida and Susan Zhang},
      year={2023},
      eprint={2304.02034},
      archivePrefix={arXiv},
      primaryClass={cs.LG},
      url={https://arxiv.org/abs/2304.02034}, 
}

@misc{coppola2025renormalizationgroupdeepneural,
      title={Renormalization group for deep neural networks: Universality of learning and scaling laws}, 
      author={Gorka Peraza Coppola and Moritz Helias and Zohar Ringel},
      year={2025},
      eprint={2510.25553},
      archivePrefix={arXiv},
      primaryClass={cond-mat.dis-nn},
      url={https://arxiv.org/abs/2510.25553}, 
}

@article{Berman_2023,
   title={Bayesian renormalization},
   volume={4},
   ISSN={2632-2153},
   url={http://dx.doi.org/10.1088/2632-2153/ad0102},
   DOI={10.1088/2632-2153/ad0102},
   number={4},
   journal={Machine Learning: Science and Technology},
   publisher={IOP Publishing},
   author={Berman, David S and Klinger, Marc S and Stapleton, Alexander G},
   year={2023},
   month=oct, pages={045011} }

@article{Berman_2025,
doi = {10.1088/2632-2153/ade04c},
url = {https://doi.org/10.1088/2632-2153/ade04c},
year = {2025},
month = {jun},
publisher = {IOP Publishing},
volume = {6},
number = {2},
pages = {025059},
author = {Berman, D S and Klinger, M S and Stapleton, A G},
title = {NCoder—a quantum field theory approach to encoding data},
journal = {Machine Learning: Science and Technology}
}

@misc{howard2024bayesianrgflowneural,
      title="{Bayesian RG Flow in Neural Network Field Theories}", 
      author={Jessica N. Howard and Marc S. Klinger and Anindita Maiti and Alexander G. Stapleton},
      year={2024},
      eprint={2405.17538},
      archivePrefix={arXiv},
      primaryClass={hep-th},
}

@article{antognini2019finite,
	title        = {Finite size corrections for neural network Gaussian processes},
	author       = {Joseph M. Antognini},
	year         = 2019,
	eprint       = {1908.10030},
	archiveprefix = {arXiv},
	primaryclass = {cs.LG}
}

@misc{hron2020infinite,
  title         = {Infinite Attention: {NNGP} and {NTK} for Deep Attention Networks},
  author        = {Hron, Jiri and Bahri, Yasaman and Sohl-Dickstein, Jascha and Novak, Roman},
  year          = {2020},
  eprint        = {2006.10540},
  archivePrefix = {arXiv},
  primaryClass  = {stat.ML}
}

@misc{halverson2024conformal,
  title        = {Conformal Fields from Neural Networks},
  author       = {James Halverson and Joydeep Naskar and Jiahua Tian},
  year         = 2024,
  eprint       = {2409.12222},
  archiveprefix = {arXiv},
  primaryclass = {hep-th}
}

@misc{ferko2025qmnn,
  title        = {Quantum Mechanics and Neural Networks},
  author       = {Christian Ferko and James Halverson},
  year         = 2025,
  eprint       = {2504.05462},
  archiveprefix = {arXiv},
  primaryclass = {hep-th}
}

@book{WessBagger1992,
  title     = {Supersymmetry and Supergravity},
  author    = {Wess, Julius and Bagger, Jonathan},
  edition   = {2nd},
  year      = {1992},
  publisher = {Princeton University Press},
  address   = {Princeton, NJ},
  series    = {Princeton Series in Physics},
  isbn      = {978-0-691-02530-9}
}

@article{McKeon_Sherry,
  title   = {Spinors and supersymmetry in four-dimensional Euclidean space},
  author  = {McKeon, D. G. C. and Sherry, T. N.},
  journal = {Annals of Physics},
  volume  = {288},
  pages   = {2--36},
  year    = {2001},
  doi     = {10.1006/aphy.2001.6114}
}

@article{Lukierski1,
  title   = {On superfield formulation of Euclidean supersymmetry},
  author  = {Lukierski, J. and Nowicki, A.},
  journal = {Journal of Mathematical Physics},
  volume  = {25},
  number  = {8},
  pages   = {2545--2549},
  year    = {1984},
  doi     = {10.1063/1.526439}
}

@article{Lukierski2,
  title   = {Euclidean supersymmetrization of instantons and self-dual monopoles},
  author  = {Lukierski, J. and Zakrzewski, W. J.},
  journal = {Physics Letters B},
  volume  = {189},
  number  = {1-2},
  pages   = {99--104},
  year    = {1987},
  doi     = {10.1016/0370-2693(87)91277-9}
}

@article{Lukierski3,
  title   = {Holomorphic and real Euclidean supersymmetries in three and four dimensions},
  author  = {Lukierski, J.},
  journal = {Czechoslovak Journal of Physics B},
  volume  = {37},
  pages   = {359--372},
  year    = {1987},
  doi     = {10.1007/BF01597262}
}

@book{PeskinSchroeder,
  author    = {Peskin, Michael E. and Schroeder, Daniel V.},
  title     = {An Introduction to Quantum Field Theory},
  series    = {Frontiers in Physics},
  year      = {1995},
  publisher = {Addison-Wesley},
  address   = {Reading, MA},
  isbn      = {978-0201503975}
}

@article{OtherPaper,
  title         = {The Neural Networks with Tensor Weights and the Corresponding Fermionic Quantum Field Theory},
  author        = {Huang, Guojun and Zhou, Kai},
  year          = {2025},
  eprint        = {2507.05303},
  archivePrefix = {arXiv},
  primaryClass  = {hep-th},
  doi           = {10.48550/arXiv.2507.05303},
}

\end{document}